\documentclass[final,authoryear,5p,times,twocolumn]{elsarticle}

\usepackage{lineno}
\usepackage{amsmath,amssymb}
\usepackage[colorlinks=True,%
            urlcolor=blue,%
            linkcolor=blue,%
            citecolor= blue,%
            pdfauthor={T. Gastine},%
            pdftitle={},%
            pdfkeywords={Atmospheres dynamics, Jupiter interior, Saturn
interior, Uranus interior, Neptune interior},%
            pdftex]{hyperref}
\usepackage{journals} 
\usepackage{booktabs}
\usepackage{longtable}

\journal{....}

\newcommand{\lp}{\left(}
\newcommand{\rp}{\right)}
\newcommand{\tb}{\tilde{T}}
\newcommand{\rb}{\tilde{\rho}}

\newcommand{\modif}[1]{\textnormal{#1}}

\def\vec#1{\ensuremath{\mathchoice{\mbox{\boldmath$\displaystyle#1$}}
{\mbox{\boldmath$\textstyle#1$}}
{\mbox{\boldmath$\scriptstyle#1$}}
{\mbox{\boldmath$\scriptscriptstyle#1$}}}}
\def\tens#1{\ensuremath{\mathsf{#1}}}

\begin{document}

\begin{frontmatter}

\title{Zonal flow scaling in rapidly-rotating compressible convection}

\author[MPS]{T.~Gastine\corref{cor1}}
\ead{gastine@mps.mpg.de}
\author[Ed]{M.~Heimpel}
\author[MPS]{J.~Wicht}

\address[MPS]{Max Planck Institut f\"ur Sonnensytemforschung, 
Justus-von-Liebig-Weg 3, 37077 G\"ottingen, Germany}
\address[Ed]{Department of Physics, University of Alberta, Edmonton, Alberta 
T6G 2J1, Canada}

\cortext[cor1]{Principal corresponding author}

\begin{abstract}
The surface winds of Jupiter and Saturn are primarily zonal. Each planet
exhibits 
strong prograde equatorial flow flanked by multiple alternating
zonal winds at higher latitudes. The depth to which these flows penetrate has
long been debated and is still an unsolved problem. Previous rotating 
convection models that obtained multiple high latitude zonal jets comparable to 
those on the giant planets assumed an incompressible (Boussinesq) fluid, which 
is unrealistic for gas giant planets. Later models of compressible rotating 
convection obtained only few high latitude jets which were not amenable to 
scaling analysis.

Here we present 3-D numerical simulations of compressible convection in
rapidly-rotating spherical shells. To explore the formation and scaling of 
high-latitude zonal jets, we consider models with a strong radial density 
variation and a range of Ekman numbers, while maintaining a zonal flow Rossby 
number characteristic of Saturn.

All of our simulations show a strong prograde equatorial jet
outside the tangent cylinder. At low Ekman numbers several alternating
jets form in each hemisphere inside the tangent cylinder. To analyse
jet scaling of our numerical models and of Jupiter and Saturn, we extend Rhines
scaling based on a topographic $\beta$-parameter, which was previously
applied to an incompressible fluid in a spherical shell,
to compressible fluids. The jet-widths predicted by
this modified Rhines length are found to be in relatively good agreement with
our numerical model results and with cloud tracking observations of
Jupiter and Saturn.
\end{abstract}


\end{frontmatter}

\section{Introduction}

The surface flows of the gas giants Jupiter and Saturn are dominated by 
strong zonal motions (i.e. azimuthal flows). Zonal wind profiles
at the surface are obtained by tracking cloud features with ground-based and 
space observations \citep[e.g.][]{Sanchez00,Porco03,Porco05,Vasavada05}. 

\begin{figure}[t]
 \centering
 \includegraphics[width=8.8cm]{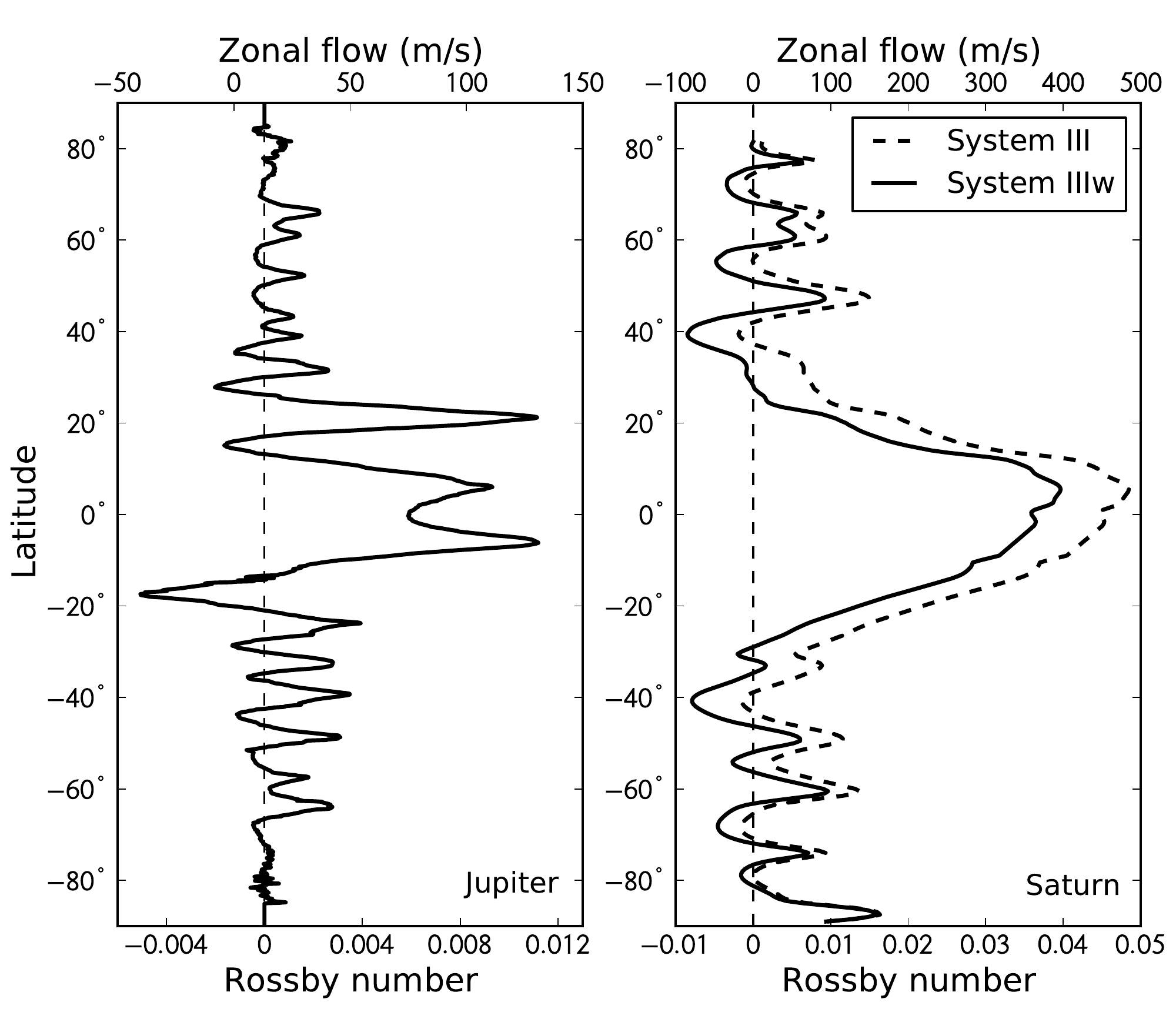}
 \caption{Surface zonal flow profiles for Jupiter (left panel) and Saturn 
(right panel). Jupiter's zonal flow profiles come from Cassini's data by 
\cite{Porco03} and \cite{Vasavada05}. Saturn's profiles in the two rotations 
systems have been derived by \cite{Read09}. The conversion between zonal winds 
velocities  in m/s and Rossby numbers is given by $u/\Omega r_o$, $r_o$ being 
the planetary radius at the $1$ bar level.}
 \label{fig:vpPlanets}
\end{figure}

As shown in Fig.~\ref{fig:vpPlanets}, these zonal winds form a differential 
rotation profile with alternating eastward and westward flows.
Both gas giants feature a strong prograde equatorial jet flanked by several 
weaker alternating secondary jets. Jupiter's central equatorial jet reaches a 
maximum velocity around  100-140 m/s and covers latitudes within 
$\pm 15^\circ$. However the region of fast equatorial zonal flow extends to 
roughly $\pm 25^\circ$, and features several strong prograde 
and retrograde jets that display marked equatorial asymmetry. The secondary 
undulating zonal winds at higher latitudes are weaker ($\sim 10-20$ m/s) and 
narrower than the equatorial flow. Saturn's equatorial flow is 
stronger than that of Jupiter, wider, and more symmetrical, with a 
single equatorial jet extending roughly $\pm 30^\circ$ in latitude. It is 
flanked by several secondary jets in each hemisphere. These winds are 
significantly shifted towards the prograde direction when assuming the rotation 
period measured by Voyager \citep{Desch81}. However, the suitability of 
this so-called System III rotation period, which is based on Saturn's 
kilometric radio (SKR) emissions, for characterizing the mean planetary
rotation rate has been questioned. The current Cassini space mission has 
measured an apparent 6 minute increase in the SKR rotation period since Voyager 
\citep[e.g.][]{Sanchez05}. This simple change of 1\% in the rotation rate can 
substantially modify the zonal wind speed (typically around 20\% near the 
equator). Since such changes in the planetary rotation rate are unlikely 
\citep{Heimpel12}, alternative rotation systems have been proposed. Here we 
adopt the System IIIw rotation period of  \cite{Read09}, which is based on an 
analysis of potential vorticity. This System IIIw rotation period 
yields more symmetric zonal flows between the prograde and the retrograde 
directions (see Fig.~\ref{fig:vpPlanets}). The amplitude of the equatorial 
flow then reduces from 450 m/s in the System III rotation period to 370 m/s in 
System IIIw.

Modelling of zonal wind dynamics can be categorised in two main 
approaches. In ``shallow models'', which are typically based on the hydrostatic 
approximation of fluid dynamics, the dynamics is confined to a 
very thin layer close to the cloud level \citep[e.g.][]{Vallis06, Liu11}. In 
this approach, zonal winds are maintained by turbulent motions coming from 
several possible physical forcings that occur at the stably-stratified cloud 
level (e.g. latent heat release or solar radiation). These shallow models 
reproduce several features observed on Jupiter and Saturn and notably the 
alternating direction of the zonal winds \citep{Williams78,Cho96}. While 
earlier models yielded a retrograde equatorial jet for both gas giants, 
prograde equatorial flow has been obtained in more recent models via the 
inclusion of additional forcing mechanisms, such as water vapor condensation 
\citep{Lian10} or enhanced radiative damping \citep[e.g.][]{Liu11}.

In an alternative approach, zonal winds may be driven by deep-seated 
convection. This is supported by the latitudinal thermal emission profiles 
of Jupiter and Saturn, which are relatively flat, implying that deep-seated, 
outward directed heat flow exceeds absorbed solar heat for both planets 
\citep{Ingersoll76, Pirraglia84}. Although deep convection must drive the 
dynamos that produce the global magnetic fields of Jupiter and Saturn, it is 
thought that fast zonal flows would be strongly attenuated at depth by the 
magnetic braking resulting from the strong increase in the electrical 
conductivity \citep{Guillot04,French12}. Based on an estimate 
of the associated Ohmic dissipation, \cite{Liu08} concluded that the zonal 
flows must therefore be confined to a relatively thin layer (i.e. $0.96\, R_J$ 
and $0.85\,R_S$). This magnetic exclusion of fast zonal flow to an outer 
region has been demonstrated in recent dynamo models that include radially 
variable electrical conductivity \citep{Heimpel11, Duarte13}.

The deep convection hypothesis has been tested by 3-D numerical models of 
turbulent convection in rapidly-rotating spherical shells 
\citep[e.g.][]{Christensen01, Heimpel05}. Rapid rotation causes convection to 
develop as axially-oriented quasi-geostrophic columns \citep{Busse70}. This 
columnar flow gives rise to Reynolds stresses, a statistical correlation between 
the convective flow components that feeds energy into zonal flows 
\citep[e.g.][]{Busse94}. The usual prograde tilt of the convective columns goes 
along with a positive flux of angular momentum away from the rotation axis that 
yields an eastward equatorial flow \citep[e.g.][]{Zhang92}. While such models 
can easily reproduce the correct direction and amplitude of the equatorial jets 
observed on Jupiter and Saturn, they usually fail to produce multiple 
high-latitude jets \citep{Christensen02}. Numerical models in relatively 
deep layers and moderately small Ekman numbers (i.e. aspect ratio $r_i/r_o=0.6$ 
and $E=10^{-4}-10^{-5}$) typically produce only a pair of jets in each 
hemisphere \citep{Christensen01,Jones09,Gastine12}. 

Reaching quasi-geostrophic turbulence in a 3-D model of rotating convection is 
indeed numerically very demanding. These numerical difficulties can be 
significantly reduced by using the quasi-geostrophic approximation which 
includes the effects of the curvature of the spherical shell (i.e. the 
topographic $\beta$-effect) while solving the flow only in the 2-D equatorial 
plane \citep{Aubert03,Schaeffer05NP}. For instance, the two-dimensional 
annulus model of rotating convection considered by \cite{Jones03,Rotvig06} and 
\cite{Teed12} allows very low Ekman numbers to be reached. Multiple 
alternating zonal flows have been found and the typical width of each jet scales 
with the Rhines length \citep[e.g.][]{Rhines75,Gurarie02,Read04,Suko07}. 
Independently of these models, the 3-D simulations of \cite{Heimpel05} and 
\cite{Heimpel07} were computed with lower Ekman numbers and larger aspect ratio 
($r_i/r_o \ge 0.85$)  than previous numerical models 
\citep[e.g.][]{Christensen01}. These simulations show clear evidence of multiple 
jets inside the tangent cylinder. While these simulations produce fewer bands 
than observed in the  quasi-geostrophic simulations, at least in part due 
to the unavoidable computational limitations, the width of each zonal band 
follows the Rhines scaling \citep[][hereafter HA07]{Heimpel07}.

Most of the previous models have employed the Boussinesq approximation where 
compressibility effects are simply ignored. In giant planets, however, the 
density \modif{increases by more than 1000 in the molecular envelope} 
\citep{Guillot99,French12} and the applicability of 
the topographic Rhines scaling is therefore questionable 
\citep{Evonuk08}. More recent anelastic models that allow to incorporate 
compressibility effects have nevertheless shown several similarities with the 
previous Boussinesq studies. While the density contrast affects the small-scale 
convective motions, the large-scale zonal flows are relatively unaffected by 
the density stratification \citep{Jones09,Gastine12}. However, these results 
seem incompatible with the new vorticity source introduced by 
compressibility, which may change the definition of the $\beta$-effect 
\citep[e.g.][]{Ingersoll82,Glatz09,Verhoeven14}. As pointed out by 
\cite{Jones09}, defining a consistent Rhines scaling in a compressible fluid is 
not straightforward.

To tackle this problem, we consider numerical models of rapidly-rotating 
convection in a compressible fluid. We present numerical simulations of a thin 
spherical shell ($r_i/r_o=0.85$) with a strong density stratification 
($\rho_\text{bot} /\rho_\text{top} \simeq 150$). A series of simulations, with 
decreasing Ekman number, are employed to gradually reach the multiple-jets 
regime. To analyse the zonal flow scaling, we present a consistent derivation 
of Rhines scaling in compressible fluids.

The paper is organised as follows. In section~\ref{sec:model}, we present 
the anelastic formulation and the numerical method. The results of the 
numerical simulations are presented in section~\ref{sec:results}. A new 
derivation of compressible Rhines scaling is presented in 
section~\ref{sec:rhines}. In section~\ref{sec:interiormodels}, we concentrate on 
the applications to numerical models and gas giants, before concluding in 
section~\ref{sec:conclusion}.

\section{Hydrodynamical model}

\label{sec:model}

\subsection{Governing equations}

We consider numerical simulations of a compressible fluid in a spherical shell 
rotating at a constant rotation rate $\Omega$ about the $z$-axis. Following 
\cite{Christensen06}, we adopt a dimensionless formulation of the Navier-Stokes 
equations where $\Omega^{-1}$ is the time unit and the spherical shell 
thickness $d=r_o-r_i$ is the reference lengthscale. Entropy is given in units 
of $\Delta s$, the fixed entropy contrast over the layer. Density and 
temperature are non-dimensionalised using $\rho_o=\rb(r=r_o)$ and 
$T_o=\tb(r=r_o)$, their reference values at the outer boundary. Heat 
capacity $c_p$, kinematic viscosity $\nu$ and thermal diffusivity $\kappa$ are 
assumed to be constant.

We employ the anelastic approximation of \cite{Brag95} and \cite{Lantz99} which 
allows to incorporate the effects of density stratification while filtering out 
fast acoustic waves \citep[see also][]{Brown12}. This approximation assumes an 
hydrostatic and adiabatic reference state given by $d\tb/dr = - g /c_p$. 
Assuming an ideal gas leads to a polytropic equation of state given by 
$\rb=\tb^m$, $m$ being the polytropic index. Following \cite{Jones09,Gastine12} 
and \cite{Gastine13}, we assume that the mass is concentrated in the inner part, 
such that $g\propto 1/r^2$ provides a good first-order approximation of the 
gravity profile in the molecular envelope of a giant planet. This leads to the 
following temperature $\tb$ and density $\rb$ profiles

\begin{equation}
  \tb(r) = \dfrac{c_0}{(1-\eta)r}+ 1-c_0 \quad\text{and}\quad\rb(r) = \tb^m,
\end{equation}
where $c_0=\eta/(1-\eta)[\exp(N_\rho/m) -1]$. $N_\rho 
=\ln \rb(r_i)/\rb(r_o)$ is the number of density scale heights of the 
background density profile and $\eta=r_i/r_o$ is the aspect ratio of the 
spherical shell.

\modif{%
The thermodynamic quantities, density, pressure and temperature are then
decomposed into the sum of the reference state and small perturbations
$\tilde{x}(r)+x'(r,\theta,\phi)$}. The dimensionless equations that govern 
compressible convection under the anelastic approximation are given by

\begin{equation}
 \vec{\nabla}\cdot \lp \rb\vec{u} \rp = 0,
 \label{eq:anel}
\end{equation}
\begin{equation}
   \dfrac{\partial \vec{u}}{\partial
   t}+\vec{u}\cdot\vec{\nabla}\vec{u}
+2\vec{e_z}\times\vec{u}
    =
 -\vec{\nabla}{\dfrac{p}{\rb}}+Ra^*\dfrac{r_o^2}{r^2} s\,\vec{e_r} 
+  \dfrac{E}{\rb} \vec{\nabla}\cdot\tens{S},
 \label{eq:NS}
\end{equation}
\begin{equation}
\rb\tb\lp\dfrac{\partial s}{\partial t} + \vec{u}\cdot\vec{\nabla} s\rp =
\dfrac{\text{E}}{\text{Pr}}\vec{\nabla}\cdot\lp\rb\tb \vec{\nabla} s\rp +
\dfrac{\text{E}}{\text{Ra}^*}(1-\eta)c_o\,Q_\nu,
\label{eq:entropy}
\end{equation}
where $\vec{u}$, $p$  and $s$ are velocity, pressure
and entropy, respectively. The traceless rate-of-strain tensor $\tens{S}$ is 
given by

\begin{equation}
\tens{S}_{ij}= 2 \rb\left(\tens{e}_{ij}- \frac{1}{3} \delta_{ij}
\vec{\nabla}\cdot\vec{u} \right) \quad\text{with}\quad
\tens{e}_{ij} = \dfrac{1}{2}\left(\frac{\partial u_i}{\partial x_j} +
\frac{\partial u_j}{\partial x_i}\right),
\label{eq:tenseur}
\end{equation}
where $\delta_{ij}$ is the identity matrix. $Q_\nu$ is the viscous heating 
contribution expressed by

\begin{equation}
 Q_\nu =
2\rb\left[\tens{e}_{ij}\tens{e}_{ji}-\dfrac{1}{3}(\vec{\nabla}\cdot\vec{u}
)^2\right ].
\end{equation}
The system of equations (\ref{eq:anel}-\ref{eq:entropy}) is 
controlled by three nondimensional numbers: the Ekman number $E=\nu/\Omega 
d^2$; the Prandtl number $Pr=\nu/\kappa$ and the modified Rayleigh number $Ra^* 
= g_o \Delta s / c_p \Omega^2 d$, $g_o$ being the gravity at the outer 
boundary. Note that $Ra^*$ relates to the conventional definition of the 
Rayleigh number via $Ra^* = RaE^2/Pr$ and that  $\sqrt{Ra^*}$ is commonly 
referred to as the convective Rossby number \citep[e.g.][]{Elliott00}.

In all the numerical models presented here, we have assumed constant entropy 
and free-slip velocity boundary conditions at both spherical shell boundaries 
$r_i$ and $r_o$.

\subsection{Numerical modelling}

\begin{table*}
\caption{\modif{Parameters of the simulations computed in this study and 
corresponding dimensionless numbers in Jupiter and Saturn.}}
\centering
\begin{tabular}{cccccc}
  \toprule
   & $E$ & $Pr$ & $Ra^*$ & $N_r\times \ell_{\text{max}} \times 
m_{s}$ & $\alpha$ \\
  \midrule
  Simulation 1 &$3 \times 10^{-4}$ &  1 & 1.71 & $97\times 288 \times 1$ & 2 \\
  Simulation 2 &$10^{-4}$ &  1 & 0.65 & $97\times 341 \times 1$ & 3\\ 
 Simulation 3 &$3 \times 10^{-5}$ &  0.3 & 0.282 & $129\times 426 \times 4$ & 
7 \\ 
 Simulation 4 &$10^{-5}$ &  0.1 & 0.18 & $145\times 682 \times 8$ & 12\\ 
 Simulation 5 &$3 \times 10^{-6}$ &  0.1 & 0.116 & $193\times 682 \times 8$& 22 
 \\ 
  \midrule
 Jupiter & $10^{-16}$ & $0.1-1$ & $2\times 10^{-4}$ & & \\ 
 Saturn & $10^{-17}$ & $0.1-1$ & $6\times 10^{-5}$ & & \\
\bottomrule
 \end{tabular}
\label{tab:runs}
\end{table*}

The numerical simulations have been computed using the anelastic version of the 
code MagIC \citep{Wicht02,Gastine12}, which has been recently
benchmarked \citep{Jones11}. The system of equations 
(\ref{eq:anel}-\ref{eq:entropy}) is solved using a poloidal-toroidal 
decomposition of the mass flux:

\begin{equation}
 \rb \vec{u} = \vec{\nabla}\times (\vec{\nabla}\times W \,\vec{e_r}) +
 \vec{\nabla} \times Z\, \vec{e_r}.
\end{equation}
The poloidal and toroidal potentials $W$ and $Z$, as well as $s$ and $p$, are 
expanded in spherical harmonic functions up to degree and order
$\ell_\text{max}$ in colatitude $\theta$ and longitude $\phi$, and in Chebyshev 
polynomials up to degree $N_r$ in radius. As indicated in Tab.~\ref{tab:runs}, 
the numerical truncations considered here range from 
($N_r=97,\,\ell_\text{max}=288$) for the case with the largest Ekman number to 
($N_r=193,\,\ell_\text{max}=682$) for the numerical model with $E=3\times 
10^{-6}$. To save some computational resources, the most demanding numerical 
simulations have been solved on an azimuthally truncated sphere with  a 
four-fold ($m_s=4$) or a eight-fold ($m_s=8$) symmetry. This enforced symmetry 
may influence the dynamics of the solution at high latitude and help the 
flow to become more axisymmetric. However, since compressible convection 
in the rapidly-rotating regime is dominated by small-scale structures, this 
assumption is not considered to have a strong effect on the solution 
\citep[e.g.][]{Christensen02,Jones09,Gastine13}.

\begin{figure}
 \centering
 \includegraphics[width=8.8cm]{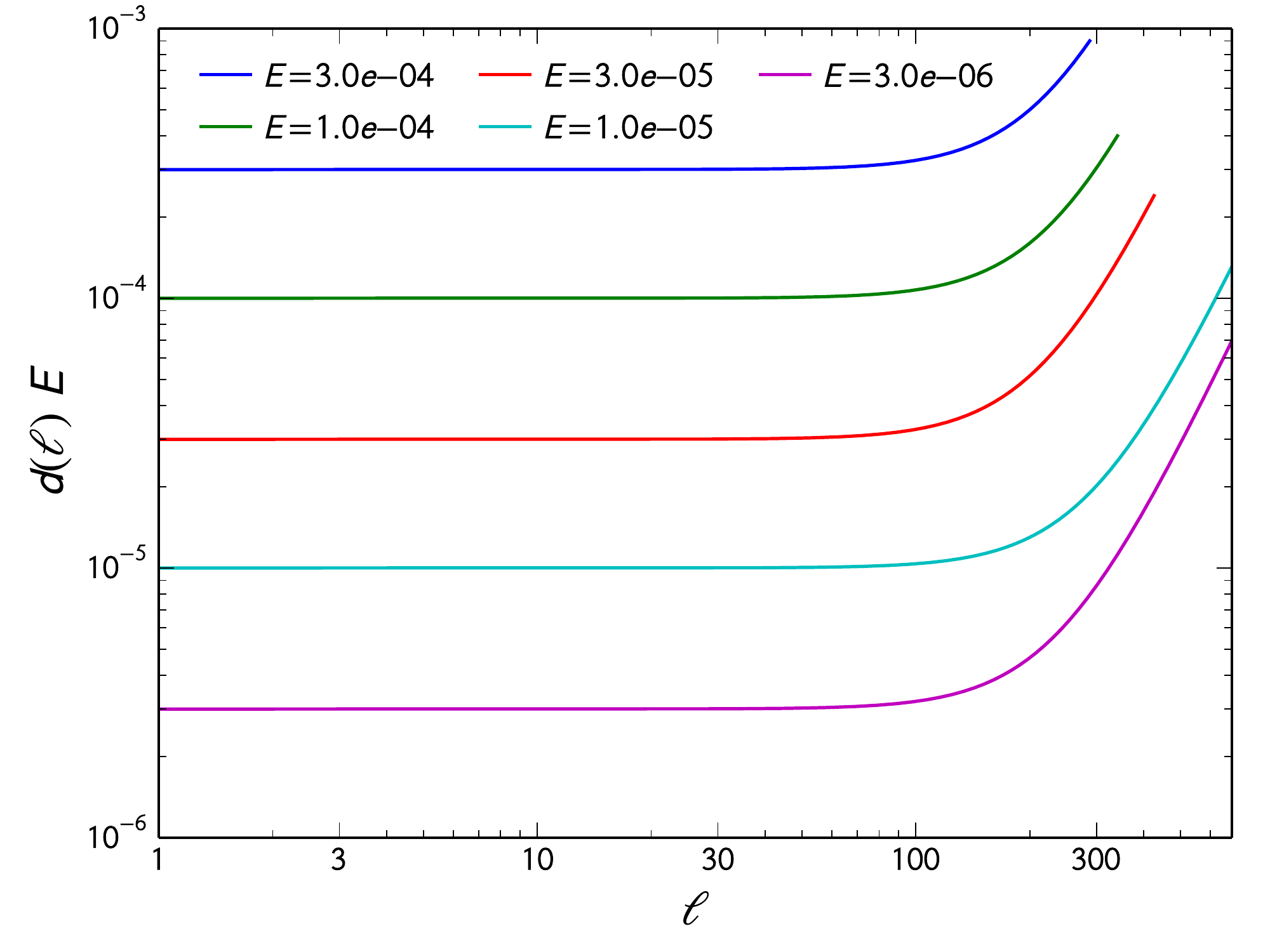}
 \caption{\modif{Hyperdiffusion profiles as a function of the spherical 
harmonic degree $\ell$ for the numerical simulations of 
Tab.~\ref{tab:runs} (Eq.~\ref{eq:hdif}).}}
 \label{fig:hdif}
\end{figure}

Despite the large grid sizes employed here, the convergence of these 
numerical models still requires the use of hyperdiffusivity. This means that 
the diffusive terms entering in Eqs.~(\ref{eq:NS}-\ref{eq:entropy}) are 
multiplied by an operator of the functional form

\begin{equation}
 d(\ell) = \lp 1+\alpha\left[\dfrac{\ell-1}{\ell_{\text{max}}-1}\right]
^\beta\rp.
 \label{eq:hdif}
\end{equation}
Here, $d(\ell)$ is the hyperdiffusivity function that depends on the spherical 
harmonic degree $\ell$, $\alpha$ is the hyperdiffusion amplitude and $\beta$ is 
the hyperdiffusion exponent \citep[e.g.][]{Kuang99}. $\beta$ is set to 3 in the 
present numerical models while $\alpha$ varies from 
$2$ for the model at $E=3\times 10^{-4}$ to $\alpha=22$ for the case at 
$E=3\times 10^{-6}$ (see Tab.~\ref{tab:runs}). \modif{Fig.~\ref{fig:hdif} shows that  
this formulation of hyperdiffusion leaves the effective molecular diffusivities 
unchanged for $\ell \lesssim 100$. For larger values of $\ell$, the diffusivities
increase to the maximum factor $(1+\alpha)$ at the truncation 
degree $\ell_{\text{max}}$.}

While ensuring the stability of numerical models at values of Ekman and 
Rayleigh numbers that would normally require higher levels of truncation, the 
use of hyperdiffusion has some potential caveats. It potentially
introduces some anisotropy between the horizontal and the radial direction as 
$d(\ell)$ depends on the horizontal scale only. In addition, hyperdiffusion 
yields an artificial viscous heating that can affect the heat transport balance
\citep{Glatz02}. However the amplitudes employed here are relatively 
weak and hyperdiffusivity seems to mainly act as a low pass-filter on 
the small-scale convective structures close to the outer boundary. As a 
reliability check, the initial amplitudes of hyperdiffusion have been 
reduced stepwise when the simulations reached a statistically 
steady-state without qualitatively affecting the results.

\subsection{Parameter choice}

The parameters of the five numerical models discussed here have been chosen to 
model the dynamics of the molecular envelope of a giant planet, within the 
accessible numerical \modif{parameter space}. Tab.~\ref{tab:runs} summarizes 
this parameter choice. 
The values of $E$ and $Ra^*$ explored here are far from estimated values for
a gas giant planet. Computational resolution limits us to a minimum Ekman 
number that is roughly 10 orders of magnitude larger than that for Jupiter. 
The Ekman number has been gradually lowered, spanning the
range from $E=3\times10^{-4}$ to $E=3\times 10^{-6}$. The values of $Ra^*$ 
have been tuned to obtain zonal \modif{Rossby number profiles of similar 
amplitudes for the different Ekman numbers.
To maintain zonal flow velocities comparable to Saturn
($Ro\simeq 0.05$ at the equator) the model $Ra^*$ values decrease with decreasing $E$. 
However, $Ra^*$ is still orders of magnitude greater than 
planetary values for the case with the lowest Ekman number. In 
contrast with our previous parameter studies, which were dedicated to the 
influence of $Ra^*$ on the zonal flow regime \citep{Gastine13,Gastine14}, here 
we consider numerical models with similar Rossby numbers and focus on the 
influence of the Ekman number.}
The Prandtl number is set to 1 for $E\ge 10^{-4}$ and then  successively 
lowered in the most demanding cases to ensure a quicker nonlinear saturation.  We 
employ a larger aspect ratio of $\eta=0.85$ than in our previous studies 
\citep{Gastine12,Gastine13} which is known to promote the formation of 
multiple high latitude jets \citep[e.g.][]{Christensen01,Heimpel05,Heimpel07}. 

The thickness of this convective layer exceeds estimates of the zonal flow depth 
of Jupiter ($\sim 0.95\,R_J$) but is comparable to estimated values for Saturn
\citep[$\sim 0.85\,R_S$, see][]{Liu08,Nettelmann13}. \modif{Using these 
estimates of the zonal flow depth as lengthscales, $\nu=3\times 
10^{-7}~\text{m}^2/\text{s}$ as 
a typical value for the kinematic viscosity in the molecular envelope and 
$\Omega_J = 1.75\times 10^{-4}~\text{s}^{-1}$ ($\Omega_S = 
1.64\times10^{-4}~\text{s}^{-1}$) allow to estimate the Ekman numbers for 
Jupiter ($E=10^{-16}$) and Saturn ($E=10^{-17}$). According to the internal 
models by \citep{French12}, the thermal diffusivity varies from $\kappa=3\times 
10^{-7}~\text{m}^2/\text{s}$ to $\kappa=2\times 10^{-6}~\text{m}^2/\text{s}$ in 
Jupiter's molecular envelope, which yields $0.15<Pr<1$. A direct estimate
of $Ra^*$ for Jupiter and Saturn is not accessible as the value of the 
superadiabatic entropy contrast $\Delta s$ is not known.} \modif{However, 
following \cite{Gastine13}, this parameter can be indirectly inferred
from the modified flux-based Rayleigh number $Ra_q^*=Ra^*\,Nu^* = \alpha g 
q/\rho c_p \Omega^3 d^2$ and the heat transport scaling law that relates the 
modified Nusselt number $Nu^*$ to $Ra_q^*$ via
$Nu^*=0.076\,(Ra_q^*)^{0.53}$ 
\citep[e.g.][]{Christensen02,Gastine12}.} \modif{The values given in 
Tab.~\ref{tab:param} then yield $Ra^{*}=2\times 10^{-4}$ and $Ra^{*}=6\times 
10^{-5}$ for Jupiter and Saturn, respectively.}

\begin{table*}
\caption{\modif{Physical properties and values assumed for Jupiter and Saturn 
\citep{Hanel81,Guillot99,French12}.}}
\centering
\begin{tabular}{ccccc}
  \toprule
   quantity & meaning (units) & Jupiter & Saturn \\
  \midrule
  $d$ & Lengthscale (m) & $0.05\,R_J=3.5\times 10^{6}$ & $0.15\,R_S=8.7\times 
10^{6}$ \\ 
  $\Omega$ & Rotation rate (s$^{-1}$) & $1.75\times 10^{-4}$ & $1.64\times 
10^{-4}$ \\
 $q$ & Heat flux (W/m$^2$) & 5.5 & 2 \\
 $\rho$ & Density (kg/m$^3$) & 250 & 330 \\
  $\alpha$ & Thermal expansivity (K$^{-1}$) & $9 \times 10^{-5}$ & $3\times 
10^{-4}$ \\
  $g$ & Gravity (m/s$^2$) & 25 & 12 \\
  $c_p$ & Heat capacity (J/kg/K) & $1.2\times 10^{4}$ & $1.2\times 10^{4}$ \\
\bottomrule
 \end{tabular}
\label{tab:param}
\end{table*}

All the simulations have a polytropic index $m=2$ and a strong density contrast 
of $N_\rho=5$, which corresponds to $\rho_{\text{bot}}/\rho_{\text{top}} \simeq 
150$. Due to the numerical limitations, this stratification is still weaker than 
the actual density contrast of the envelope of a giant planet: for instance, 
Saturn's model by \cite{Nettelmann13} suggests $N_\rho=7.4$ between $0.85\,R_S$ 
and the 1 bar level \citep[see also][]{Guillot99}. The most demanding setup 
explored here is similar to the case F computed by \cite{Jones09}, albeit
with a smaller Ekman number and stronger zonal flows ($Ro\sim 0.05$ compared 
to $Ro\sim 0.01$)\footnote{Note that there is $(1-\eta)^2$ factor difference in 
the Rayleigh number definition adopted by \cite{Jones09} compared to our 
definition.}

The numerical models with $E\ge 10^{-4}$ have been initiated from a random 
entropy perturbation superimposed on the conductive thermal state. The 
converged solutions of these runs have then been used as a starting condition 
for the more demanding cases, the Ekman number being lowered in stages. 
All the cases were run for at least 0.5 viscous diffusion time, ensuring that a 
statistical steady state has been reached.

\section{Results of the numerical models}

\label{sec:results}

\begin{figure*}
 \centering
 \includegraphics[width=18cm]{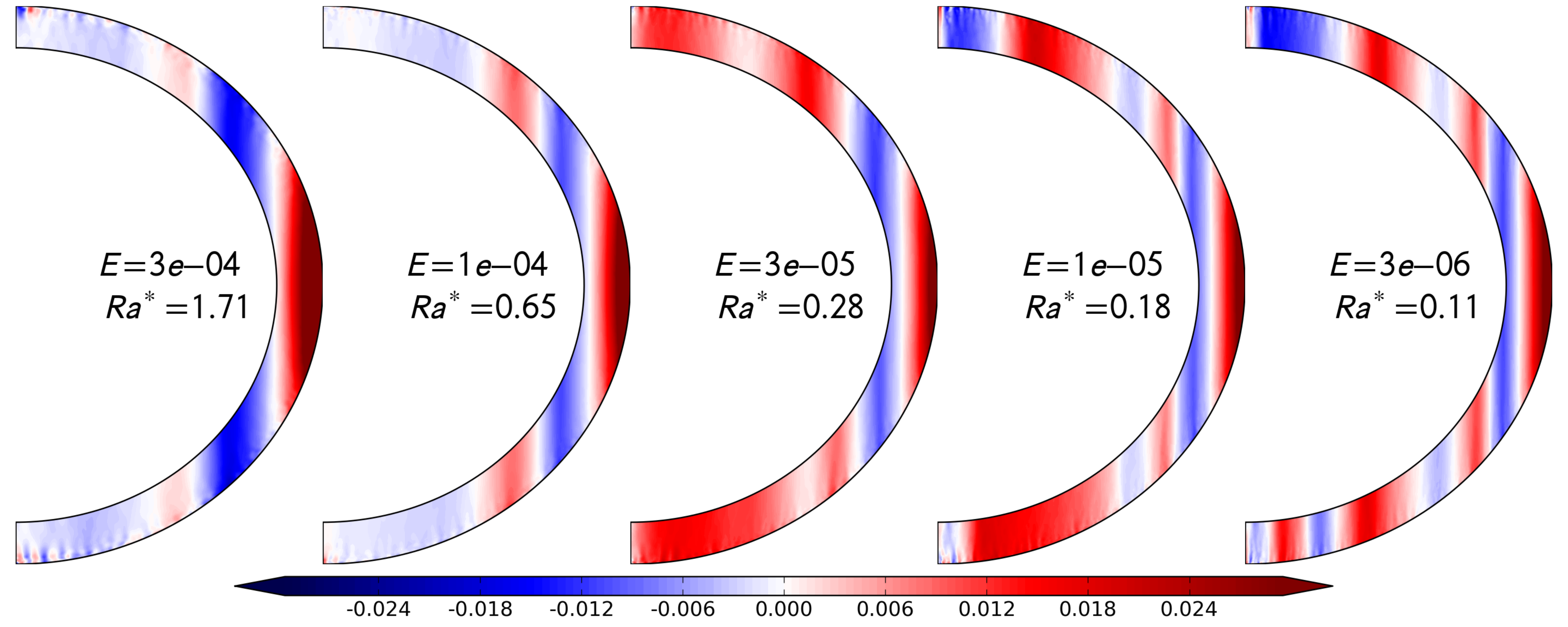}
 \caption{\modif{Snapshots of} zonally averaged azimuthal velocity in the 
meridian plane for the numerical models of Tab.~\ref{tab:runs}. Velocities are 
expressed in Rossby number units (i.e. $u_\phi/\Omega r_o$) and colorscales are 
centered around zero: prograde (retrograde) jets are rendered in red (blue). In 
all cases, the prograde contours have been truncated in amplitude to emphasize 
the structure of the secondary jets.}
 \label{fig:vps}
\end{figure*}

\begin{figure*}
 \centering
 \includegraphics[width=18cm]{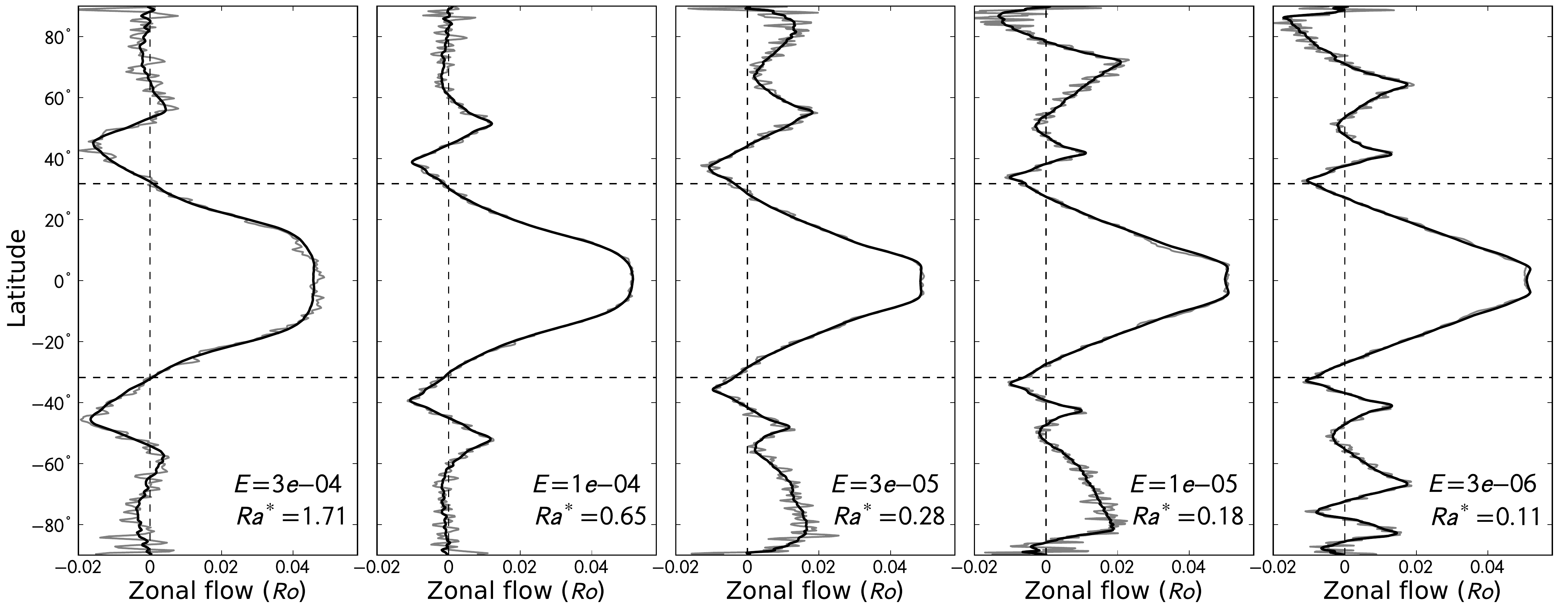}
 \caption{Surface zonal flow profiles of the numerical models of 
Tab.~\ref{tab:runs} (expressed in units of $Ro = u_\phi / \Omega r_o$). The 
grey lines correspond to the snapshots displayed in Fig.~\ref{fig:vps}, while 
the black lines correspond to time-averaged zonal flows. On each panel, the 
location of the tangent cylinder is marked by the two horizontal dashed lines.}
 \label{fig:vpsurface}
\end{figure*}

Fig.~\ref{fig:vps} shows the zonal flow structure for the five Ekman numbers 
employed here. The values of $Ra^*$ are set to keep the Rossby number at an 
approximately constant value. Fig.~\ref{fig:vpsurface} portrays the 
corresponding latitudinal profiles of the surface zonal flows. For $E\ge 
10^{-4}$ (the two first panels), the main prograde equatorial jet is flanked by 
retrograde zonal flows attached to the tangent cylinder. At mid-latitudes (i.e. 
$\theta \simeq \pm 50-60^\circ$) a pair of weak prograde jets develops in a 
similar way to our previous thicker-shell simulations \citep{Gastine12}. Typical 
of strongly stratified numerical models, the amplitude of the main equatorial 
jet is significantly larger than those of the secondary jets ($Ro\simeq 0.05$ at 
the equator compared to $Ro \simeq 0.01$ at $\theta\pm 45^\circ$). While the 
zonal winds are dominated by geostrophic flows, some ageostrophic features are 
discernable at high latitudes, especially near the outer boundary. These 
small-scale structures are strongly time-dependent and live on a typical 
convective turnover timescale. These features are responsible 
for the time-dependence visible in the surface zonal flow 
profiles (grey lines in Fig.~\ref{fig:vpsurface}) but disappear when 
time-averaged zonal flows are considered (black lines in 
Fig.~\ref{fig:vpsurface}). \modif{The geostrophic zonal flow profiles are 
consistent with the results of previous anelastic models by \cite{Jones09} and 
\cite{Gastine12}. \cite{Kaspi09} and \cite{Showman10}
developed a modified anelastic approach that neglects viscous heating but 
includes a more realistic equation of state. In their models at $E\simeq 
10^{-4}$, the zonal flows show a significant variation in the direction of the 
rotation axis, in contrast with our findings \citep[and][]{Jones09,Gastine12}. A 
possible explanation for this difference might come from the 
different heating mode employed in the different models. While, following 
\cite{Jones09}, we adopt here a fixed entropy contrast, \cite{Kaspi09} used 
flux-boundary conditions and internal heating. This yields significant 
latitudinal entropy contrasts which promotes baroclinic and ageostrophic zonal 
flows via the thermal-wind balance:}

\begin{equation}
 2\dfrac{\partial u_\phi}{\partial z} \simeq Ra^* \dfrac{r_o^2}{r^3} 
\dfrac{\partial s}{\partial \theta}.
\end{equation}
\modif{In contrast, the latitudinal variation of entropy remains weak in our 
model leading to the nearly perfectly geostrophic zonal flows ($\partial u_\phi 
/\partial z \simeq 0$) 
shown in Fig.~\ref{fig:vps}}.

For the intermediate cases with $E=3\times10^{-5}$ and $E=10^{-5}$ (third and 
fourth panels), additional jets start to appear at mid and high latitudes. The 
secondary jets already visible in the previous cases with $E\ge 10^{-4}$ 
are gradually squeezed towards lower latitudes. Some of the simulations shown 
here have a higher number of jets in \modif{one hemisphere} (here the southern 
hemisphere). As 
reported by \cite{Jones09}, this symmetry breaking is quite common in the 
multiple-jets regime and is sensitive to the initial conditions
used in the numerical models (for another example of such an asymmetry, 
see the $\eta=0.9$ case by HA07). At $E=3\times 10^{-6}$ (last panels), the 
multiple jets inside the tangent cylinder are well-delimited and reach an 
amplitude of $Ro\simeq 0.015-0.02$. The number of jets is still smaller 
than observed on Saturn. This may be due to the model Ekman number which, even 
at the computationally challenging value of $E=3\times 10^{-6}$, is still 
orders of magnitude greater than estimates of $E$ for giant planets 
(\modif{Tab.~\ref{tab:runs}}).  
Nevertheless, it is evident from Fig.~\ref{fig:vpsurface} that the jet 
wavelength of our models is approaching an asymptotic value with decreasing 
Ekman number. Thus, we find that the latitudinal extent of the zonal flows and 
the relative amplitude of equatorial to  secondary jets shows some similarities 
with Saturn's surface winds (see Fig.~\ref{fig:vpPlanets}).

\section{Zonal flow scaling in compressible convection}

\label{sec:rhines}

\subsection{Jet scaling in quasi-geostrophic turbulence}

In a typical three-dimensional turbulent flow, the energy directly cascades 
from large-scale eddies to smaller-scale structures. However, due to the 
dominance of the Coriolis force, rapidly-rotating flows are strongly 
anisotropic. The turbulent structures are aligned with the rotation axis and 
form a quasi-geostrophic flow which approximately satisfies the 
Taylor-Proudman theorem. This quasi-two-dimensionalisation of the flow promotes 
the development of an inverse energy cascade which transfers energy from 
small-scale to larger-scale structures \citep[e.g.][]{Suko07}. 
In addition, due to the latitude-dependent effect of the Coriolis force, 
the flows can also be characterized by Rossby waves 
with wavenumbers that depend on the so-called $\beta$-effect. In a 
2-D shallow layer model (under the $\beta$-plane approximation), the 
$\beta$-effect can be attributed to the latitudinal variations of the Coriolis 
parameter $f=2\Omega \sin \theta$, leading to $\beta=1/r_o\,(df/d\theta)$ 
where $\theta$ is the latitude and $r_o$ is the radius of the 2D spherical 
surface \citep[e.g.][]{Williams78}. Alternatively, in a quasi-geostrophic flow 
that develops in a 3D spherical shell, the \textit{topographic} $\beta$-effect 
depends on the relative height variation of the shell with the distance to the 
rotation axis: $\beta =2\Omega\,(d\ln h /ds)$, where $s$ is the cylindrical 
radius and $h$ is the height of the spherical shell 
\citep[e.g.][HA07]{Aubert03,Schaeffer05NP}. In his original derivation, 
\cite{Rhines75} proposed that the inverse cascade ceases at a typical 
lengthscale defined by $L_\beta \sim k_\beta^{-1}\sim 
(2U_\text{rms}/\beta)^{1/2}$, subsequently referred to as the Rhines scale. In 
other words, $k_\beta$ separates the spectrum on turbulence (smaller scales with 
$k> k_\beta$) and Rossby waves (larger scales with $k<k_\beta$). Furthermore, 
$k_\beta$ is related to the zonation of the turbulent flow which forms 
alternating jets of typical width $k_\beta^{-1}$. The interpretation of 
the Rhines wavelength is however still debated and the precise role 
played by the inverse cascade on the zonation mechanism remains controversial 
\citep{Suko07,Tobias11,Sri12}. Despite these uncertainties, this lengthscale 
plays a major role in many natural systems encompassing oceanic circulation 
\citep[e.g.][]{Vallis93}; atmosphere dynamics \citep[e.g.][]{Held96,Schneider04} 
and circulation in the gas giants atmospheres \citep[][HA07]{Vasavada05}.

In a compressible fluid, the exact definition of the $\beta$ parameter 
that enters the Rhines wavelength is, however, not clear. Indeed, 
\cite{Evonuk08} and \cite{Glatz09} pointed out that the compressibility adds a 
new vorticity source that could potentially alter the definition of $L_\beta$. 
In their two-dimensional numerical models, the flow is cylindrical, with 
no boundary curvature effect (i.e. $\beta=0$) and the density only varies in a 
direction perpendicular to the rotation axis. In this particular setup, 
$L_\beta$ can thus be simply replaced by $L_\rho = (2 
U_\text{rms}/\beta_\rho)^{1/2}$, with $\beta_\rho = 
2\Omega\,(d\ln\rb/ds)$, $s$ being the cylindrical radius \citep[see 
also][]{Verhoeven14}. In a 3-D spherical 
shell, however, this compressional Rhines scale is unlikely to be directly 
applicable. First, the height of the container varies with the 
distance to the rotation axis and the classical topographic $\beta$ will 
therefore play a role. Secondly, the density varies with the spherical radius 
$r$ and the definition of $L_\rho$ based on $s$ must take this into 
account \citep{Jones09}.

\subsection{\texorpdfstring{$\beta$-effect in a compressible 
fluid}{beta-effect in a compressible fluid}}
A formulation of the $\beta$-parameter that depends on the variation of 
the axially integrated mass with distance from the rotation axis was proposed by 
\cite{Ingersoll82} \modif{for quasi-geostrophic flows}. Here we verify this 
scaling argument by deriving the compressional $\beta$-parameter and Rhines 
scale in a spherical shell in a consistent way with the momentum 
equation~(\ref{eq:NS}). To do this we first examine the flow structure in the 
compressible quasi-geostrophic limit. Taking the curl of Eq.~(\ref{eq:NS}) gives 
the following vorticity equation

\begin{equation} 
\begin{aligned}
\dfrac{D\vec{\omega}}{Dt}+\vec{\omega}(\vec{\nabla}\cdot\vec{u})-(\vec{\omega}
\cdot\vec{\nabla})\vec{u} & =  2\left[\dfrac{\partial\vec{u}}{\partial z} 
-(\vec{\nabla}\cdot \vec{u})\vec{e_z} \right] \\
  & +\vec{\nabla}\times \lp Ra^*\dfrac{r_o^2}{r^2}s\,\vec{e_r }
+\dfrac{E}{\rb} \vec{\nabla}\cdot\tens{S}\rp,
\end{aligned}
\label{eq:vorticity}
\end{equation}
where $D/Dt$ corresponds to the substantial time derivative. In the 
limit of 
small Ekman and Rossby numbers, the first-order contribution of this equation 
comes from the Coriolis force. Retaining only the first term on the right hand 
side of Eq.~(\ref{eq:vorticity}), and using the continuity equation 
(\ref{eq:anel}), in cylindrical coordinates $(s,\phi,z)$, one thus obtains 
\modif{in the geostrophic limit}

\begin{equation}
 \dfrac{\partial u_s}{\partial z} = \dfrac{\partial u_\phi}{\partial z} = 0
\quad;\quad \dfrac{\partial u_z}{\partial z} = -\dfrac{d \ln\rb}{dr} u_r.
\label{eq:TP}
\end{equation}
This is the Taylor-Proudman theorem reformulated for a compressible fluid 
\citep[see also][]{Tortorella,Jones09}. $u_\phi$ and $u_s$ are
$z$-independent as in the Boussinesq approximation, while $u_z$ depends on $z$, 
owing to the strong density gradients close to the surface. 
The first-order flow contributions can thus be expressed as $u_s(s,\phi)$, 
$u_\phi(s,\phi)$ and $u_z(s,\phi,z)$. This also means that the $z$-vorticity 
component (i.e. $\omega_z = (\vec{\nabla}\times u)_z$) is a function of $s$ and 
$\phi$ only \modif{in the geostrophic limit.}

To derive the compressible formulation of the $\beta$-effect, we focus on the 
$z$-component of Eq.~(\ref{eq:vorticity}). \modif{Following \cite{Rhines75}, we 
first establish the dispersion relation for pure compressible Rossby waves, 
using the linearised $z$-vorticity equation without source and dissipation 
terms}. We consider in the following the 
dimensional formulation of this equation to ease further comparisons 
with observations.

\begin{equation}
  \dfrac{\partial \omega_z}{\partial t} = 2\Omega \lp\dfrac{\partial 
u_z}{\partial z} + \vec{u}\cdot\vec{\nabla}\ln\rb\rp.
\label{eq:vortPure}
\end{equation}
Expanding the density gradient in cylindrical coordinates then leads to

\begin{equation}
  \dfrac{\partial \omega_z}{\partial t} = 2\Omega \lp\dfrac{\partial 
u_z}{\partial z} + \dfrac{1}{\rb}\dfrac{\partial \rb}{\partial s}u_s
+\dfrac{1}{\rb}\dfrac{\partial \rb}{\partial z}u_z
\rp.
\end{equation}
We then multiply this equation by $\rb$ and average over the 
height $h$ of the spherical shell, where $h = 2\sqrt{\smash{r_o^2}-s^2}$ outside 
the tangent cylinder and $h=\sqrt{\smash{r_o^2}-s^2}-\sqrt{\smash{r_i^2}-s^2}$ 
inside

\begin{equation}
   \dfrac{1}{h}\int_h \dfrac{\partial \rb {\omega}_z}{\partial t} dz = 2\Omega 
\lp \dfrac{1}{h}\int_h \dfrac{\partial \rb u_z}{\partial z}dz
+ \dfrac{1}{h}\int_h \dfrac{\partial \rb}{\partial s} u_s dz \rp.
\end{equation}
Using the anelastic Taylor-Proudman theorem given in Eq.~(\ref{eq:TP}), $u_s$ 
and $\omega_z$ can be taken out of the integrals, leading to

\begin{equation}
 \dfrac{M}{h} \dfrac{\partial \omega_z}{\partial t} = 2\Omega \lp
\underbrace{\dfrac{1}{h}\int_h \dfrac{\partial \rb u_z}{\partial z}dz}_{(a)}
+ \underbrace{\dfrac{u_s}{h} \int_h \dfrac{\partial \rb}{\partial 
s}dz}_{(b)} \rp,
\label{eq:eq1}
\end{equation}
where $M=\int_h \rb dz$ is the integrated mass along the axial direction.
The first integral in the right hand side of this equation is a surface 
integral that can be simplified using the impermeable flow boundary 
condition $u_r[r_i,r_o]=0$. In cylindrical coordinates, this leads to

\begin{equation}
 \cos\theta\, u_z(r=[r_i,r_o]) +\sin\theta\, u_s(r=[r_i,r_o]) = 0.
\end{equation}
Using $\cos \theta = z/r$ and $\sin \theta =s/r$ then leads to

\begin{equation}
 (a) = \left\lbrace
\begin{aligned}
 & -\dfrac{\rho_o s}{r_o^2-s^2}\bar{u}_s  \,\text{for}\, s \geq r_i, \\
& \dfrac{s\left(\rho_i\sqrt{\frac{r_o^2-s^2}{r_i^2-s^2}}-\rho_o 
\sqrt{\frac{r_i^2-s^2}{r_o^2-s^2}}\,\right)}{\sqrt{r_o^2-s^2}-\sqrt{r_i^2-s^2}} 
\bar{u}_s  \,\text{for}\, s < r_i,
\end{aligned}
\right.
\label{eq:part1}
\end{equation}
where $\rho_o$ and $\rho_i$ are the density at outer and inner radii, 
respectively. The second integral of Eq.~(\ref{eq:eq1}) can be rearranged using 
the Leibniz rule of differentiation under the integral sign:

\begin{equation}
 \displaystyle \dfrac{\partial M}{\partial s} =\dfrac{\partial }{\partial s} 
\int_h \rb 
dz = \left[\rb\dfrac{d h}{ds} \right]_h + \int_h \dfrac{\partial \rb}{\partial 
s} dz.
\end{equation}
One thus gets

\begin{equation}
 (b)=\left\lbrace
\begin{aligned}
&\dfrac{1}{h}\dfrac{\partial M}{\partial s}\bar{u}_s + \dfrac{\rho_o 
s}{r_o^2-s^2}\bar{u}_s \,\text{for}\, s \geq r_i, \\
&\dfrac{1}{h}\dfrac{\partial M}{\partial s}\bar{u}_s 
- \dfrac{s\left(\rho_i\sqrt{\frac{r_o^2-s^2}{r_i^2-s^2}}-\rho_o 
\sqrt{\frac{r_i^2-s^2}{r_o^2-s^2}}\,\right)}{\sqrt{r_o^2-s^2}-\sqrt{r_i^2-s^2}} 
\bar{u}_s \,\text{for}\, s < r_i.
\end{aligned}
\right.
\label{eq:part2}
\end{equation}
\modif{The sum of $(a)$ and $(b)$ simply yields}

 \begin{equation}
\dfrac{\partial \omega_z}{\partial t} =  \beta_\rho u_s 
\quad\text{with}\quad
 \beta_\rho = \dfrac{2\Omega }{M} 
  \dfrac{\partial M}{\partial s}.
 \label{eq:betanel}
 \end{equation}
In a compressible fluid, the $\beta$-effect therefore depends on the variation 
of the $z$-integrated mass with the distance to the rotation axis. This 
confirms the expression suggested by \cite{Ingersoll82} for a 
\modif{quasi-geostrophic} anelastic fluid. In the Boussinesq limit, 
$\beta_\rho$ reduces to the topographic $\beta$ effect derived by HA07:

\begin{equation}
\left\lbrace
\begin{aligned}
 \dfrac{\beta_h}{2\Omega} = \dfrac{1}{h}\dfrac{dh}{ds}  & = 
-\dfrac{s}{r_o^2-s^2}; \quad s \geq r_i, \\
 \dfrac{\beta_h}{2\Omega} = \dfrac{1}{h}\dfrac{dh}{ds}  & = 
\dfrac{s}{\sqrt{\smash{r_o^2}-s^2}\sqrt{\smash{r_i^2}-s^2}} 
;\quad 
s < r_i. \\
\end{aligned}\right.
\label{eq:topoBeta}
\end{equation}

\begin{figure}
 \centering
 \includegraphics[width=8.8cm]{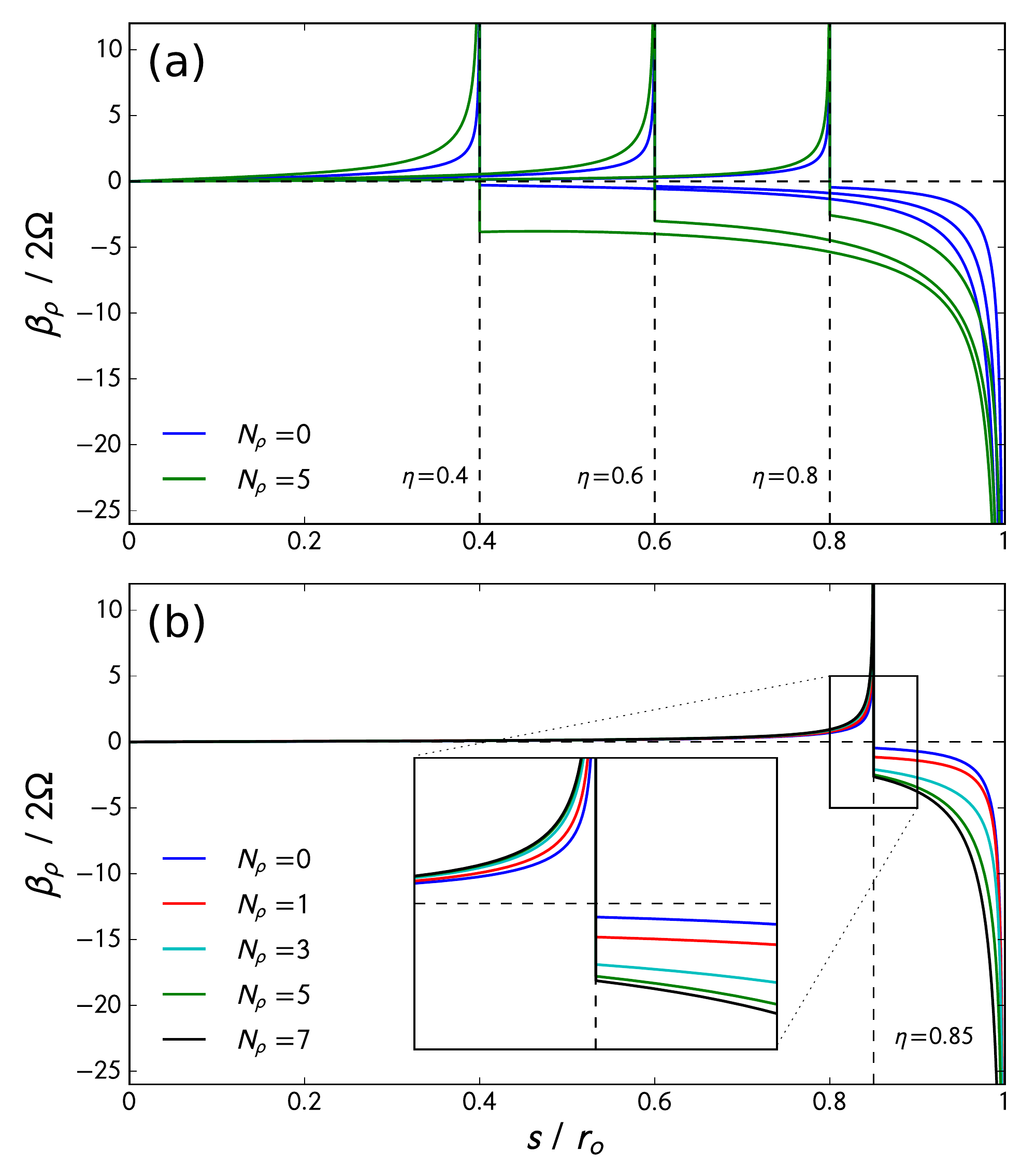}
 \caption{Upper panel: $\beta_\rho$ profiles for different spherical 
shell geometries $\eta=[0.4, 0.6, 0.8]$ with $N_\rho=0$ and $N_\rho=5$. Lower 
panel: $\beta_\rho$ profiles for different density stratifications 
$N_\rho=[0,1,3,5,7]$ with $\eta=0.85$. On each panel, the location of the 
tangent cylinder is marked by a vertical dashed line.}
 \label{fig:beta}
\end{figure}

Figure~\ref{fig:beta} shows some examples of such $\beta_\rho$ profiles for 
different spherical shell geometries and density stratifications. Outside the 
tangent cylinder, the integrated mass decreases when going away from the 
rotation axis resulting in negative $\beta_\rho$. The opposite 
happens inside the tangent cylinder. The differences with the classical 
topographic $\beta_h$ profiles (blue curves in Fig.~\ref{fig:beta}a) are more 
pronounced in the case of a relatively thick spherical shell ($\eta=0.4$ or 
$\eta=0.6$). For a thinner shell ($\eta=0.8$), this difference fades away for 
$s<r_i$ but $\beta_\rho$ is still significantly larger in amplitude 
than $\beta_h$ outside the tangent cylinder due to the strong density gradient 
in the equatorial region. This is highlighted in the lower panel of 
Fig.~\ref{fig:beta} where we consider several density contrasts in a thin 
spherical shell with $\eta=0.85$. In such a thin shell, $\beta_\rho$ increases 
regularly with $N_\rho$ outside the tangent cylinder. Inside, the profiles are 
nearly unaffected by the density stratification except in the immediate vicinity 
of the tangent cylinder.

\subsection{Compressible Rhines scale}

The Rhines scale is obtained when the phase velocity of Rossby waves
equals the mean rms flow velocity. We thus need to determine the 
dispersion relation of the anelastic Rossby wave. With the compressible 
Taylor-Proudman theorem given in Eq.~(\ref{eq:TP}), the 
continuity equation $\vec{\nabla}\cdot \rb \vec{u}=0$ applied to the 
first-order flow can be rewritten as

\begin{equation}
 \dfrac{1}{s}\dfrac{\partial s u_s}{\partial s} + 
\dfrac{1}{s}\dfrac{\partial u_\phi}{\partial \phi} =0.
\end{equation}
Introducing a streamfunction $\psi$, $u_s$ and $u_\phi$ are expressed as

\begin{equation}
 u_s = \dfrac{1}{s}\dfrac{\partial \psi}{\partial \phi} 
\quad\text{and}\quad
 u_\phi = -\dfrac{\partial \psi}{\partial s}.
\end{equation}
Noting that $\omega_z = -\Delta \psi$, the linear inviscid $z$-vorticity 
equation (Eq.~\ref{eq:betanel}) can be rewritten as

\begin{equation}
 \dfrac{\partial \Delta \psi}{\partial t} = 
-\dfrac{\beta_\rho}{s}\dfrac{\partial \psi}{\partial \phi}.
\end{equation}
Looking for plane waves with $\psi = \hat{\psi} \exp i(\vec{k}\cdot 
\vec{r}-\omega t)$, we obtain the following dispersion relation

\begin{equation}
 \omega = -\dfrac{\beta_\rho k_\phi}{k^2},
\label{eq:rossbywavedisp}
\end{equation}
where $k^2 = k_s^2+k_\phi^2$. This is identical to the dispersion 
relation of \cite{Rhines75}, except that $\beta_\rho$ replaces Rhines' 
$\beta$-plane parameter for 2-D turbulence. Due to the sign change of 
$\beta_\rho$ at the tangent cylinder (see Fig.~\ref{fig:beta}), the anelastic 
Rossby waves propagate westward inside the tangent cylinder and eastward outside
in a similar way as the classical Rossby waves.

\modif{Retaining buoyancy driving and viscosity in the vorticity equation 
(\ref{eq:vortPure}) would lead to a set of coupled differential equations. 
There 
is no easy way to derive the exact dispersion relation of the thermal Rossby 
waves in spherical geometry with $\rb$, $\tb$ and $s$ varying with radius. A 
possible workaround to establish an approximate dispersion relation would be to 
consider an anelastic annulus in which the convective forcing is maintained in 
the $s$ direction. In addition, the diffusion operators are more complicated 
than a simple Laplacian in an anelastic fluid as they involve partial 
derivatives of $\rb$. As a first-order approximation and in analogy with the 
Busse annulus, we can nevertheless speculate that an additional factor that 
involves the Prandtl number might enter the dispersion relation of the thermal 
Rossby waves \citep[e.g.][]{Busse70}. An approximate dispersion relation for 
\emph{compressible thermal Rossby waves} would then be
\begin{equation}
 \omega_{th} \sim -\dfrac{\beta_\rho k_\phi}{(1+Pr)\,k^2}.
\label{eq:dispthrossby}
\end{equation}
Comparing with Eq.~(\ref{eq:rossbywavedisp}) we see that a 
correction factor to the classical Rossby waves is therefore 
moderate for $Pr= 0.1-1$, which is relevant for the gas giants. Owing to the 
uncertainties associated with the derivation of Eq.~(\ref{eq:dispthrossby}) and 
the rather small correction factor, we will therefore work in the following 
with the dispersion relation of classical Rossby waves 
(Eq.~\ref{eq:rossbywavedisp}).}

Following \cite{Rhines75}, the phase velocity (given by $v_p=\Re(\omega/k)$)
is obtained by taking an  average orientation of the wave crests. This 
leads to $v_p \sim \beta_\rho / 2 k^2$. Equating $v_p$ with $U_\text{rms}$ 
yields the compressible Rhines wavenumber $k_\beta$ given by

\begin{equation}
 k_\beta = \sqrt{\dfrac{|\beta_\rho |}{2U_\text{rms}}} = 
\sqrt{\dfrac{\Omega}{U_\text{rms}} \left|\dfrac{\partial \ln 
M}{\partial s}\right|}.
\label{eq:rwn}
\end{equation}
From that, the Rhines wavelength in the $s$ direction is given by

\begin{equation}
 \lambda_g = \dfrac{2\pi}{k_\beta}.
  \label{eq:rwl}
\end{equation}
Note that HA07 used the Rhines wavelength in the $\theta$ direction 
to ease the comparison with the planetary observations. Here we employ 
the cylindrical coordinates which are more natural in the quasi-geostrophic 
limit.

\section{Applications of the compressional Rhines scaling}
\label{sec:interiormodels}

\subsection{An application to a numerical model}

\begin{figure}
 \centering
 \includegraphics[width=7cm]{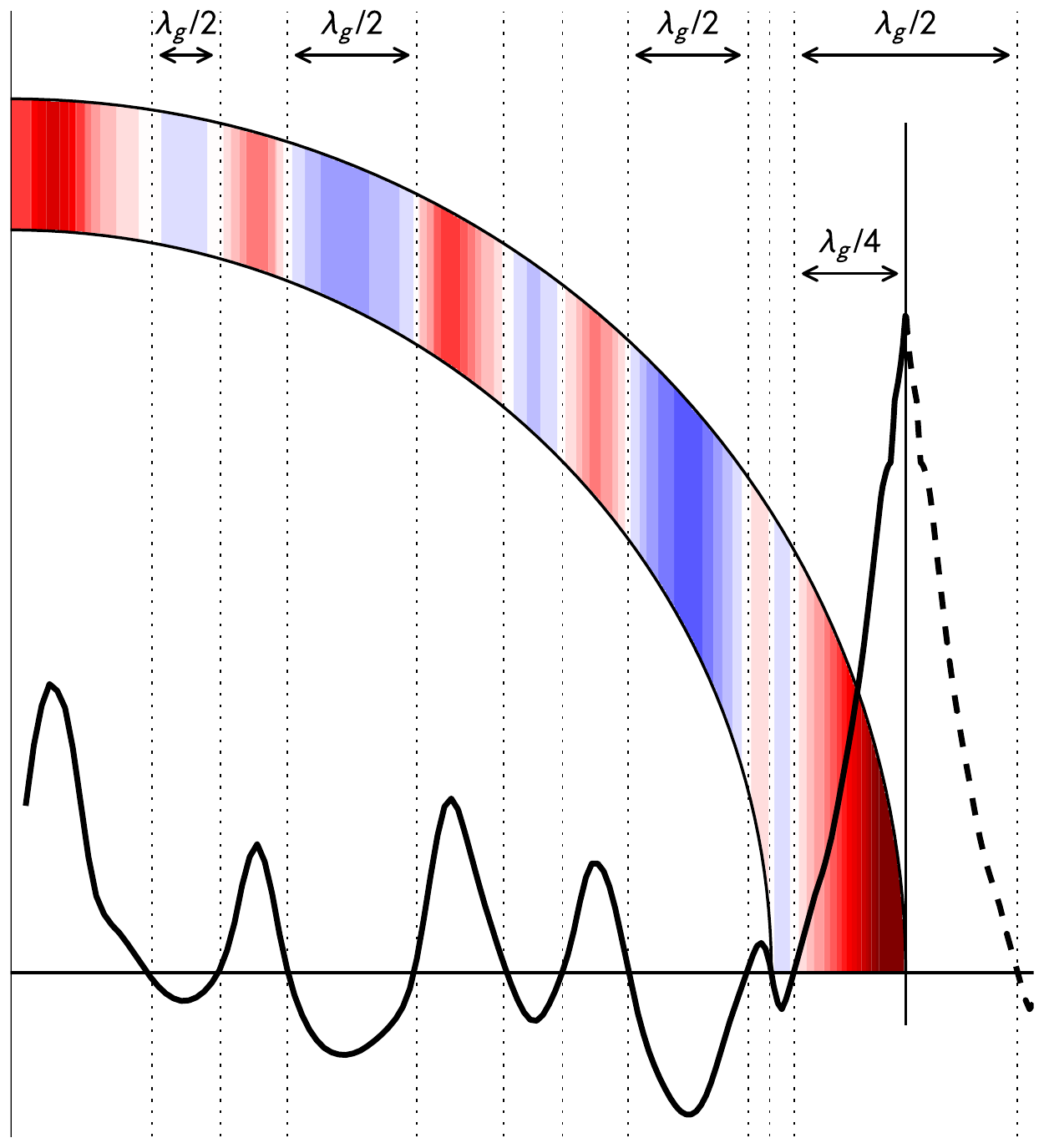}
 \caption{Sketch explaining the definition of the jets boundaries used in 
this study. \modif{The solid black line corresponds to an arbitrary surface 
zonal flow profile plotted against the cylindrical radius $s$, while the dotted 
vertical lines show the zonal flow boundaries.} $\lambda_g$ refers to the Rhines 
wavelength defined in Eq.~(\ref{eq:rwl}).}
 \label{fig:sketch}
\end{figure}

To check the influence of the compressional $\beta$-effect, we test the 
applicability of the Rhines scaling to our numerical results. \modif{%
As visible on Figs.~\ref{fig:vps} and \ref{fig:vpsurface}, the different 
numerical models have comparable Rossby numbers. As these models also share the 
same reference density profile, the Rhines scaling would therefore predict the 
same number of jets in these different models according to Eq.~(\ref{eq:rwn}). 
In contrast, we observe a gradual increase of the number of jets when the Ekman 
number is decreased (Fig.~\ref{fig:vps}). Although this might challenge the 
applicability of the Rhines scaling to our numerical models, there are some 
encouraging hints that suggest that the jet wavelength in our model at 
$E=3\times 10^{-6}$ is approaching an asymptotic state. Indeed, the zonal bands 
at low and mid latitudes (i.e. $|\theta| < 45^\circ$) have already a very 
similar structure between the $E=10^{-5}$ and the $E=3\times 10^{-6}$ cases and 
only the high-latitude jets continue to further develop when decreasing the 
Ekman number. This indicates that our models are converging towards the 
saturated state of the jet wavelength suggested by the Rhines scaling for a 
given value of $U_{\text{rms}}/\Omega$. As a consequence we focus here on the 
model with the smallest Ekman number  ($E=3\times 10^{-6}$).} 

The jet width is defined in cylindrical coordinates as the distance between two 
successive minima of $|\bar{u}_\phi|$. Here, the overbar corresponds to an 
azimuthal average. A jet boundary can therefore occur either at a sign change 
of $\bar{u}_\phi$ or, more rarely, \modif{at a local minimum}. As shown in 
Fig.~\ref{fig:sketch}, this measured jet width corresponds to half the Rhines 
wavelength $\lambda_g/2$. Considering symmetry about the equator we take the 
outer boundary ($s=r_o$) as the center of the equatorial jet, as shown in 
Fig.~\ref{fig:sketch}.

To evaluate the jet width predicted by the Rhines scaling, the right hand side 
of Eq.~(\ref{eq:rwl}) is averaged for each jet separately. Strictly speaking 
the total rms velocity should be considered when evaluating $k_\beta$. However,
for comparison with the gas giants, where the zonal velocity component 
dominates, consideration of $\bar{u}_\phi$ is sufficient. In addition, we note 
that for our simulation at $E=3\times 10^{-6}$, the energy budget is dominated 
by the zonal flow contribution ($E_\text{zon}/E_\text{tot} \simeq 0.97$),
so that using $\bar{u}_\phi$ does not significantly affect the results.

\begin{figure}
 \centering
 \includegraphics[width=8.8cm]{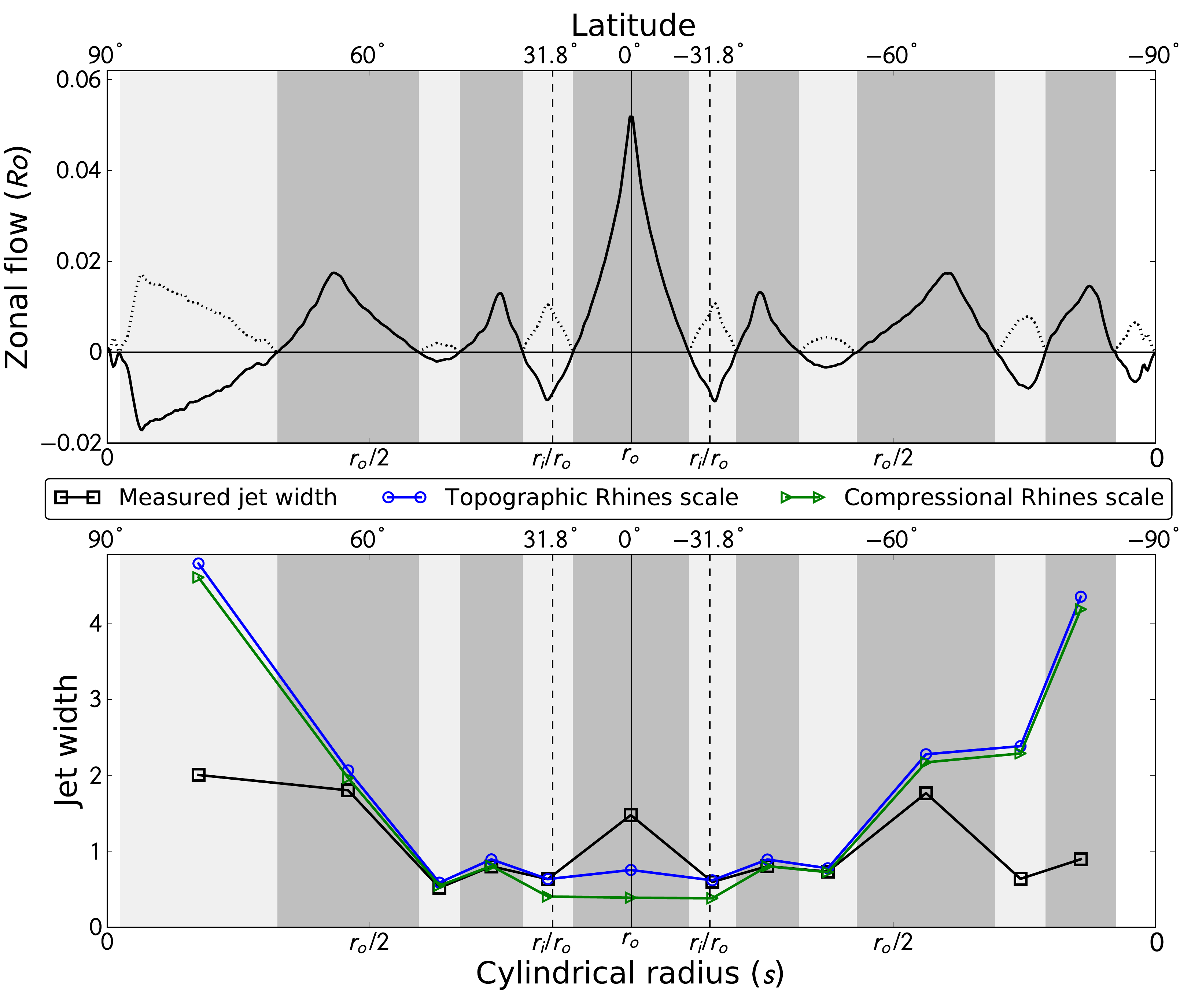}
 \caption{Upper panel: surface zonal flow profiles plotted against the 
cylindrical radius $s$ for a numerical model with $E=3\times 10^{-6}$, 
$N_\rho=5$, $\eta=0.85$, $m=2$ (last panels of 
Figs.~\ref{fig:vps}-\ref{fig:vpsurface}). Lower panel: comparison of measured 
and predicted jet width for this numerical model. The jets boundaries (defined 
as on Fig.~\ref{fig:sketch}) are successively highlighted by light and dark 
grey shaded area. The location of the tangent cylinder is marked by vertical 
dashed lines: $r_i/r_o=0.85$ corresponds to a latitude of $31.8^\circ$.}
\label{fig:vpModel}
\end{figure}

The upper panel of Fig.~\ref{fig:vpModel} shows the surface zonal flow profile 
of this numerical model as a function of $s$. This simulation has 
several jets inside the tangent cylinder which are highlighted by different grey 
\modif{shading}. The lower panel of Fig.~\ref{fig:vpModel} shows the 
comparison between 
these measured zonal flow widths and the theoretical values based on the Rhines 
scale. Inside the tangent cylinder, due to the similarities between 
$\beta_\rho$ and $\beta_h$ (see Fig.~\ref{fig:beta}b), topographic and 
compressional Rhines scale give very similar predictions. Both profiles are in 
relatively good agreement with the observed widths of the mid-latitude jets 
(i.e. $r_o/2<s<r_i/r_o$ or $30^\circ<|\theta|<60^\circ$). As 
$\beta_\rho$ tends to zero at the poles, the Rhines wavelength becomes 
very large close to the rotation axis and fails to predict the 
observed width of the high-latitude jets ($s<r_o/2$ or $\theta > 60^\circ$).
Outside the tangent cylinder, the compressional Rhines 
wavelength predicts a narrower jet as the topographic one, owing to 
$|\beta_\rho| > |\beta_h|$ there. Both underestimate the width of the 
equatorial jet by at least a factor of 2. This confirms previous 
observations by HA07 which have shown that the Rhines scale does not 
apply at the equator \citep[see also][]{Williams78}. The width of 
the equatorial jet seems instead to be largely controlled by the \modif{shell 
gap} \citep[e.g.][]{Christensen01,Heimpel05,Gastine12}.

\subsection{An application to giant planets}

\begin{figure}
 \centering
 \includegraphics[width=8.8cm]{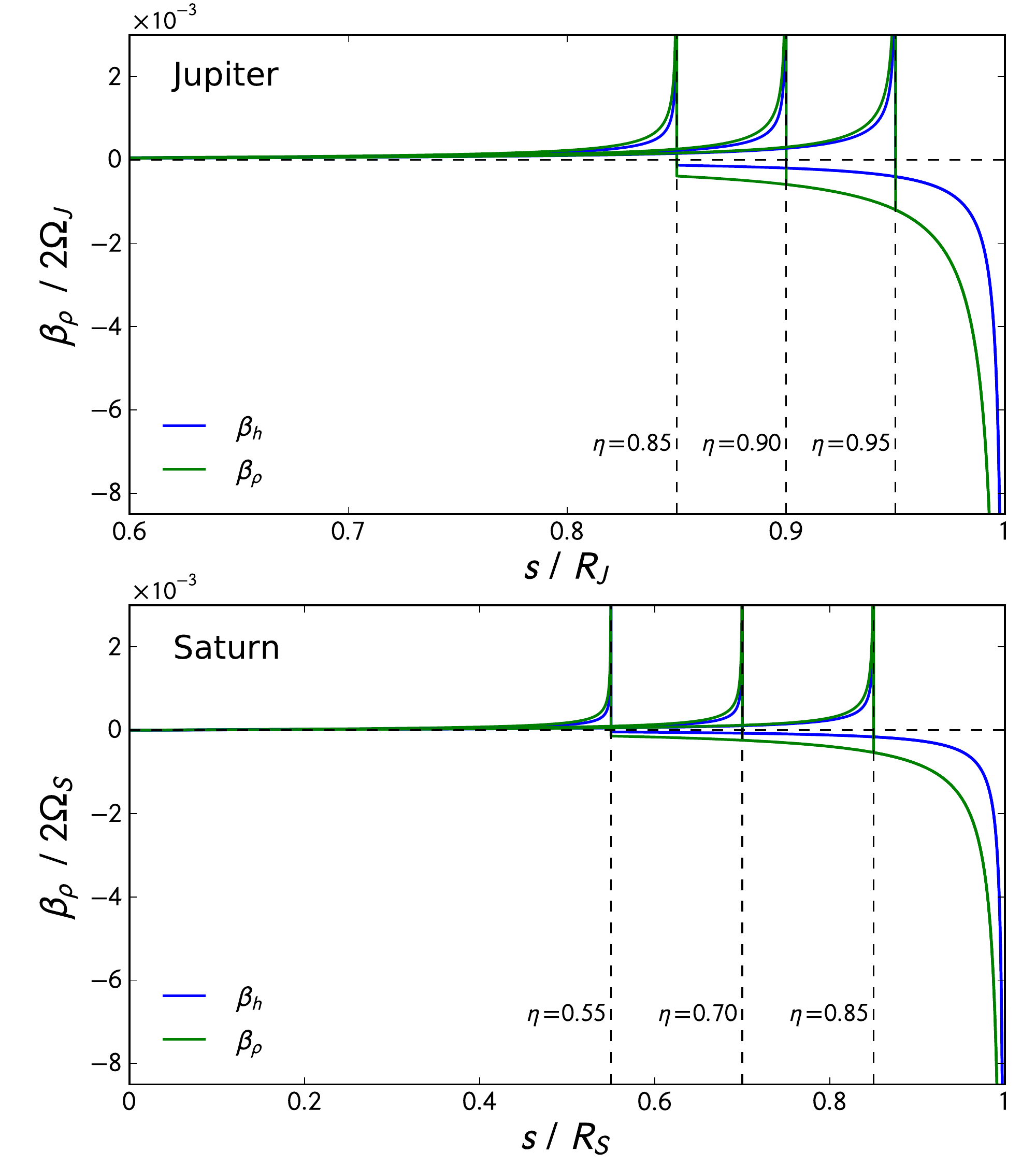}
 \caption{$\beta_\rho$ and $\beta_h$ profiles for Jupiter and Saturn 
\citep[calculated from][]{French12,Nettelmann13}. \modif{The vertical dashed 
lines correspond to different possible locations of the tangent cylinder}.}
 \label{fig:betaPlanets}
\end{figure}

\begin{figure*}
 \centering
 \includegraphics[width=15cm]{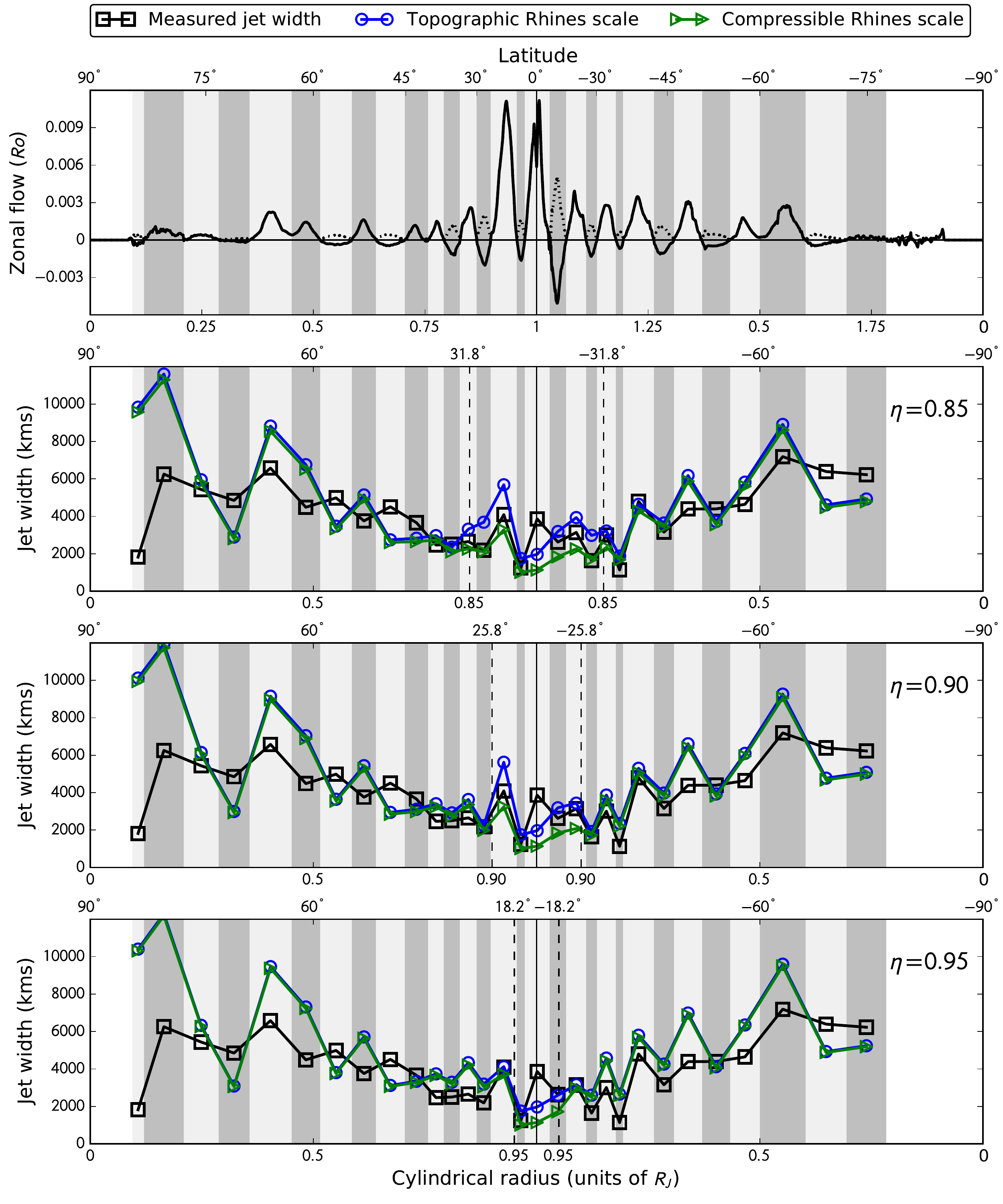}
 \caption{Upper panel: Jupiter's zonal flow profiles from \cite{Porco03}. 
Lower panels: comparison of measured and predicted jet widths for three 
different aspect ratios: $\eta=0.85$, $\eta=0.9$ and $\eta=0.95$. The jet
boundaries (defined as on Fig.~\ref{fig:sketch}) are successively highlighted 
by light and dark grey shaded area. \modif{The dashed line in the upper panel 
corresponds to the absolute value of the surface zonal flows.} The locations of 
the tangent cylinders are marked by vertical dashed lines.}
\label{fig:jupRhines}
\end{figure*}

\begin{figure*}
 \centering
 \includegraphics[width=15cm]{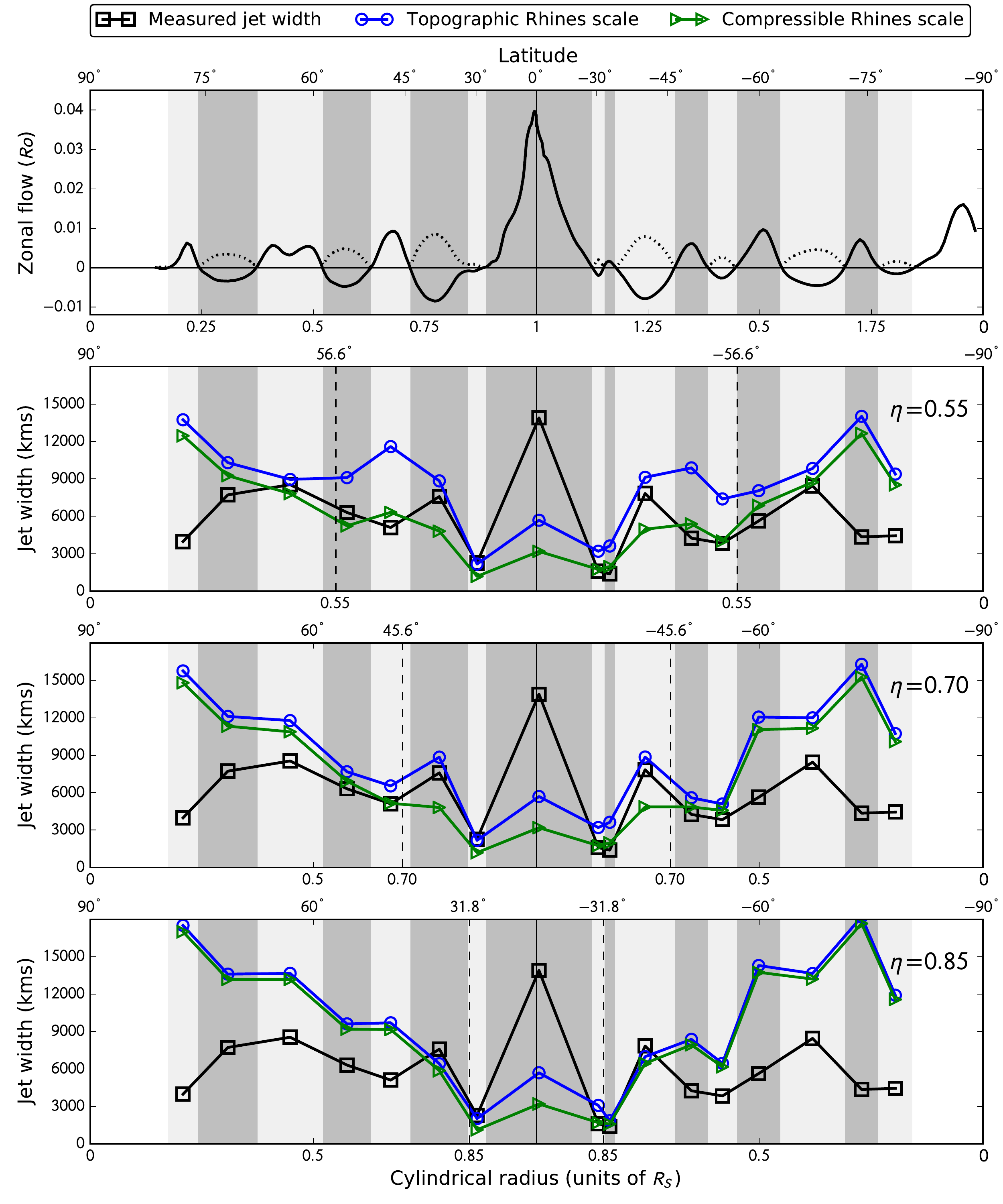}
 \caption{Upper panel: Saturn's zonal flow profiles from \cite{Read09}. Lower 
panels: comparison of measured and predicted jet widths for three different 
aspect ratios $\eta=0.55$, $\eta=0.7$ and $\eta=0.85$. The jet boundaries 
(defined as on Fig.~\ref{fig:sketch}) are successively highlighted by light and 
dark grey shaded area. \modif{The dashed line in the upper panel 
corresponds to the absolute value of the surface zonal flows.} The locations of 
the tangent cylinders are marked by vertical dashed lines.}
\label{fig:satRhines}
\end{figure*}

\modif{The deep structure of the zonal jets in the gas giant 
interiors is still a matter of ongoing debates. The recent models by 
\cite{Kaspi09}, for instance, suggest an important variation of the zonal flow 
amplitude along the axis of rotation which is not captured by our numerical 
simulations. It is therefore not clear if the compressional Rhines scale derived 
within the quasi-geostrophic limit is still applicable if the zonal flows show 
such a pronounced $z$-dependence. Keeping this limitation in mind, we 
nevertheless test the applicability of the compressional Rhines scaling to 
giant planets.}

To estimate the width of the zonal jets, we use the observed surface zonal flow 
profiles by \cite{Porco03} for Jupiter and  by \cite{Read09} for Saturn (see 
Fig.~\ref{fig:vpPlanets}). Figure~\ref{fig:betaPlanets} shows 
the radial distribution of $\beta_\rho$ determined from the 1-D interior models 
by \cite{French12} and \cite{Nettelmann13}. These profiles \modif{are 
consistent with the strongly stratified cases (i.e. $N\rho \geq 5$) 
displayed in Fig.~\ref{fig:beta}}.

Following the same procedure as for the numerical model, 
we compare on Figs.~\ref{fig:jupRhines}-\ref{fig:satRhines} the measured jet 
width with the predicted values derived from the Rhines scaling 
(Eq.~\ref{eq:rwl}). Owing to the uncertainties on the depth of the zonal jets 
in giant planets, we consider several possible locations of the tangent 
cylinder. These values have been chosen to cover the actual estimated zonal flow 
depth, ranging from the entire molecular layer (i.e. $0.55\,R_S$ and 
$0.85\,R_J$) to the shallower layers predicted by \cite{Liu08} (i.e. $0.85R_S$ 
and $0.95\,R_J$).

For Jupiter (Fig.~\ref{fig:jupRhines}), the influence of the compressibility on 
$k_\beta$ is difficult to disentangle from the topographic Rhines scaling except 
in the small region close to the tangent cylinder. In a similar way as for the 
numerical model, both $\beta_\rho$ and $\beta_h$ predict a much narrower 
equatorial jet than observed. The widths of the low-latitude  off-equatorial 
jets (i.e. $10^\circ < |\theta| < 30^\circ$) are in general better predicted by 
the compressional Rhines scale than by the incompressible one. The scaling of 
these low-latitude jets is independent of the thickness of the layer. At 
mid-latitudes ($30^\circ<|\theta|<60^\circ$), both profiles are generally 
in a good agreement with the observations. We nevertheless observe an 
increasing misfit at these latitudes when shallower layers are considered 
(visible here for the $\eta=0.95$ case). At higher latitudes ($|\theta| > 
60^\circ$), the misfit becomes larger and this can be attributed to the 
singularity of the Rhines scale when $|\theta|$ tends toward $90^\circ$.
The previous study by HA07 based on the topographic 
$\beta$-effect showed that for Jupiter the best-fit is achieved when the 
tangent cylinder is located at $0.86\,R_J$. We find a similar result, with 
predicted values in good agreement with the observations when $\eta 
\simeq 0.85-0.9\,R_J$. However, the improvement of the fit in that 
range, compared to the shallower cases, is relatively small. It is therefore 
difficult to derive a strong constraint on the zonal flow depth based on Rhines 
scaling.

For Saturn (Fig.~\ref{fig:satRhines}) the differences between calculated and
observed values are, on average, larger. The prediction based on 
$\beta_\rho$ is in better agreement with the observations than $\beta_h$, 
except for the equatorial jet which is much narrower in the compressible case.
Excluding this equatorial jet, the best-fit is obtained for the thicker layer 
with $\eta=0.55$. Considering a thinner shell indeed leads to much 
broader mid and high latitude (i.e. $|\theta| > 45^\circ$) jets than observed.

On the other hand, if one focuses on the scaling of the equatorial jet 
only, the constraint on the depth of the molecular layer is quite different. 
In fact, in the numerical models at low Ekman numbers (i.e. $E < 
1\times 10^{-5}$), the latitudinal extent of the equatorial jet is always 
controlled by the geometry of the spherical shell in such a way that the first 
retrograde jet is attached to the tangent cylinder \citep[see 
Fig.~\ref{fig:vpsurface} and][]{Christensen01,Heimpel05,Jones09,Gastine12}. For 
example, when $r_i/r_o=0.85$, the first minimum is located at $\arccos r_i/r_o 
\simeq 31.8^\circ$. Applying this geometrical argument to giant planets would 
lead to a transition radius around $0.96\,R_J$ and $0.86\,R_S$, in 
relatively close agreement with the values derived by \cite{Liu08}. This again 
underlines the difficulties to put a strong constraint on the depth of the zonal 
winds based on the Rhines scaling.

\subsection{Asymptotic scaling laws}

\modif{Using the compressional Rhines scaling also allows to construct an 
asymptotic scaling law for the number of zonal jets. This number is estimated 
as the ratio of the outer radius $r_o$ to the mean Rhines length 
$\langle\lambda_g\rangle$:
\begin{equation}
 n_{\text{jets}} = 2\dfrac{r_o}{\langle\lambda_g\rangle} = 
\dfrac{r_o}{\pi}\sqrt{\dfrac{\Omega\,d}{U_{\text{rms}}}} 
\sqrt{\dfrac{1}{d}\left\langle\left| \dfrac{\partial \ln M}{\partial 
s}\right|\right\rangle}.
\label{eq:jetscale}
\end{equation}
The factor $2$ comes from the summation of the two hemispheres, and the 
triangular brackets denote an average value of the relative mass variation 
$|\partial \ln M/\partial s|$. The jet scaling thus directly 
involves the inverse square root of the Rossby number $Ro = U_{rms}/\Omega d$.} 

\begin{figure}
 \centering
 \includegraphics[width=8.8cm]{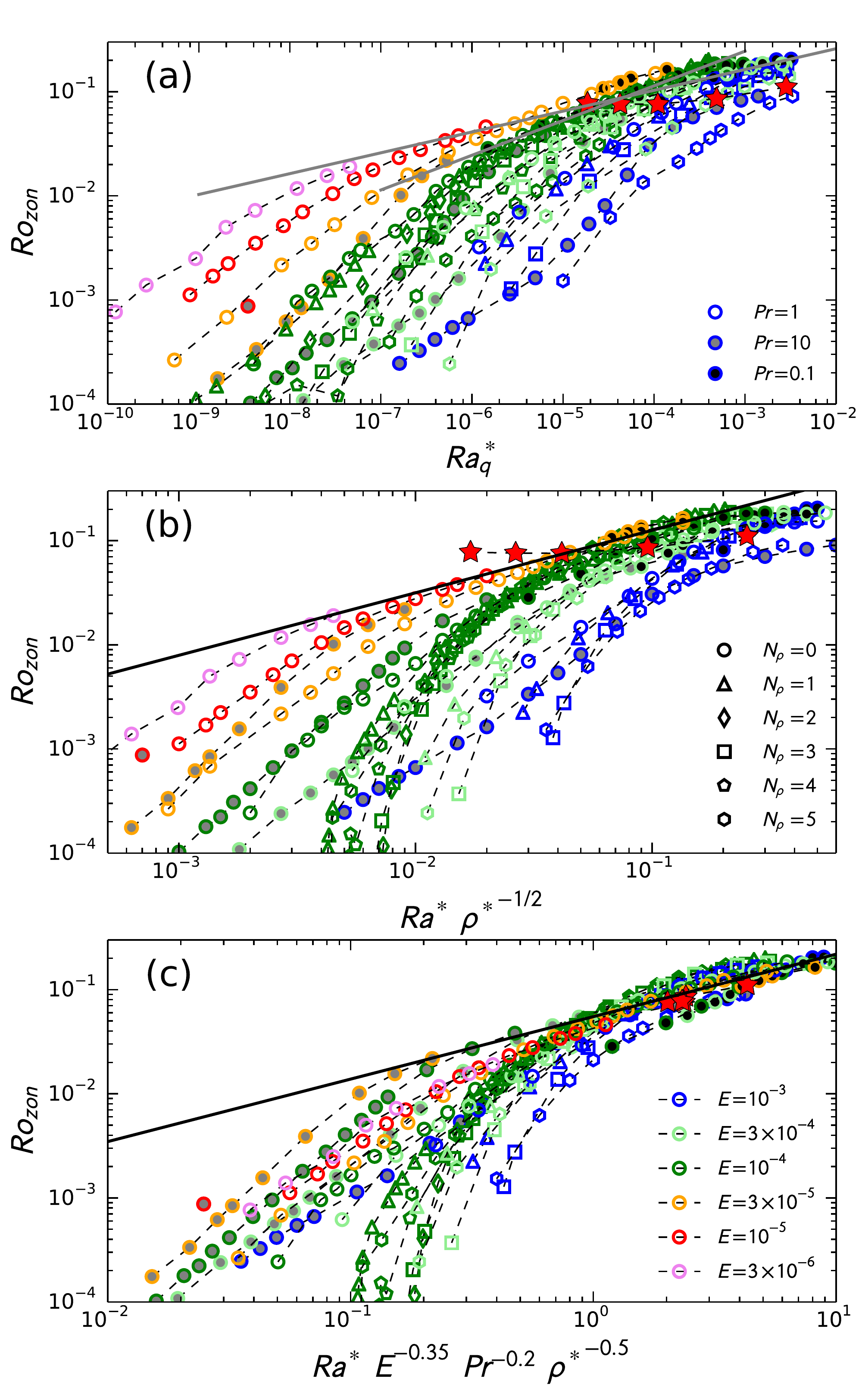}
 \caption{\modif{Different possible scaling laws for the zonal Rossby number 
$Ro_{\text{zon}}$ using the Boussinesq and the anelastic numerical models by 
\cite{Gastine12,Gastine13} and \cite{Gastine14}. \textbf{a)} $Ro_{\text{zon}}$ 
versus the modified flux-based Rayleigh number. The two solid grey lines 
correspond to the scaling derived by \cite{Christensen02} and by 
\cite{Gastine13}. \textbf{b)} $Ro_{\text{zon}}$ versus the modified Rayleigh 
number $Ra^*$. \textbf{c)} $Ro_{\text{zon}}$ versus 
$Ra^*\,E^{-0.35}\,Pr^{-0.2}\,{\rho^*}^{-0.5}$. The solid black lines in the 
panels b) and c) correspond to the asymptotic scaling laws. In each panel, the 
five red stars correspond to the numerical simulations of Tab.~\ref{tab:runs}. 
For the three panels, the color of the rim of the symbols corresponds to the 
Ekman number, the inner color to the Prandtl number and the shape to the density 
contrast.}}
 \label{fig:scaling}
\end{figure}

\modif{Several attempts have recently been made to derive an asymptotic 
scaling for the zonal Rossby number using 3-D numerical models of 
rapidly-rotating convection in spherical shells. 
Using Boussinesq models, \cite{Christensen02} suggested that 
the zonal Rossby number depends on the available buoyancy-power 
(given by the flux-based Rayleigh number $Ra_q^*$) and obeys 
the following scaling law:
\begin{equation}
 Ro_{\text{zon}} = \dfrac{U_{\text{zon}}}{\Omega d}  = 
\alpha\,\left(Ra_q^*\right)^\beta,
\end{equation}
where $(\alpha=0.65,\,\beta=1/5)$. More recently, \cite{Showman10} and 
\cite{Gastine12} derived similar convective power scaling laws using 
anelastic models of rotating convection. However, the exponents vary from
$(\alpha=1,\,\beta=1/4)$ in \cite{Showman10} 
to $(\alpha=2.45,\,\beta=1/3)$ in \cite{Gastine13}. The first panel of
Fig.~\ref{fig:scaling} shows an example of such a power-based scaling applied 
to the 
large database of Boussinesq and anelastic models of convection in 
rapidly-rotating spherical shells built in our previous parameter 
studies \citep{Gastine12,Gastine13,Gastine14}. Using a
spherical shell with an aspect ratio $\eta=0.6$, this database spans the range 
of $ 3\times 10^{-6} < E < 10^{-3}$, $Pr\in[0.1, 1, 10]$ and $0<N_\rho<5$.
As described by \cite{Showman10}, the $Ro_{\text{zon}}$ scaling undergoes 
several 
slope changes that may be attributed to some successive regime transitions. For 
moderate driving of convection, $Ro_{\text{zon}}$ increases relatively fast 
with $Ra_q^*$. The slope then gradually tapers off and seems, at first 
glance, to saturate at a level compatible with the $1/5$ scaling suggested by 
\cite{Christensen02} (upper gray line). However, a closer look indicates that 
the Ekman number dependence is not well-captured as the different lines 
associated with the different Ekman numbers are nearly parallel, each of them 
following a scaling close to $1/3$ (lower gray line). In addition, the 
impression of an asymptotic scaling is 
also artificially reinforced by the $x$-axis spanning 8 decades of 
$Ra_q^*$ \citep{Yadav13}.}

\modif{Alternatively, \cite{Christensen02} also suggested that 
$Ro_{\text{zon}}$ might directly scale as a function of the modified Rayleigh 
number $Ra^*$, a proxy of the ratio between buoyancy and Coriolis force in the 
global force balance:
\begin{equation}
 Ro_{\text{zon}} = \alpha\, (Ra^*)^\beta.
\end{equation}
Fig.~\ref{fig:scaling}b shows such a scaling applied to the database of 
Boussinesq and anelastic models. The prefactor $\alpha$ shows some dependence 
on the background density contrast, which can be easily captured 
when introducting the normalised density $\rho^*$ as suggested by 
\cite{Yadav13a}:
\begin{equation}
 \rho^* =\dfrac{1}{V}\int_v \rb\,dV.
\end{equation}
Using $(\rho^*)^{-1/2}$ then allows to nicely collapse the different 
density stratifications for one given Ekman number (open green symbols for 
instance). Although the scatter is significantly reduced compared to the 
previous $Ra_q^*$ scaling, defining an asymptotic state remains relatively 
awkward. Buoyancy indeed starts to play a significant role in the force balance 
when $Ra^*$ approaches unity. This change in the force balance is accompanied 
by a sharp transition in the zonal flow regime when $Ra^* \gtrsim 1$.
\citep{Aurnou07,Gastine13,Gastine14}.
This is already visible on the right part of Fig.~\ref{fig:scaling}b where we 
observe a gradual flattening of the $Ro_{\text{zon}}$ scaling. We can 
nevertheless bound the rapidly-rotating regime to the numerical models that 
fulfill $Ra^*\,(\rho^*)^{-1/2}<0.1$. This yields the tentative asymptotic 
scaling $Ro_{\text{zon}}=0.5\,(Ra^*/{\rho^*}^{1/2})^{0.6}$ shown by the 
solid black line in Fig.~\ref{fig:scaling}b. The thin shell simulations of 
the present study, visible as the five red stars, are however relatively far 
from the asymptotic scaling, suggesting a possible additional dependence on the 
geometry of the spherical shell.}

\begin{figure}
 \centering
\includegraphics[width=8.8cm]{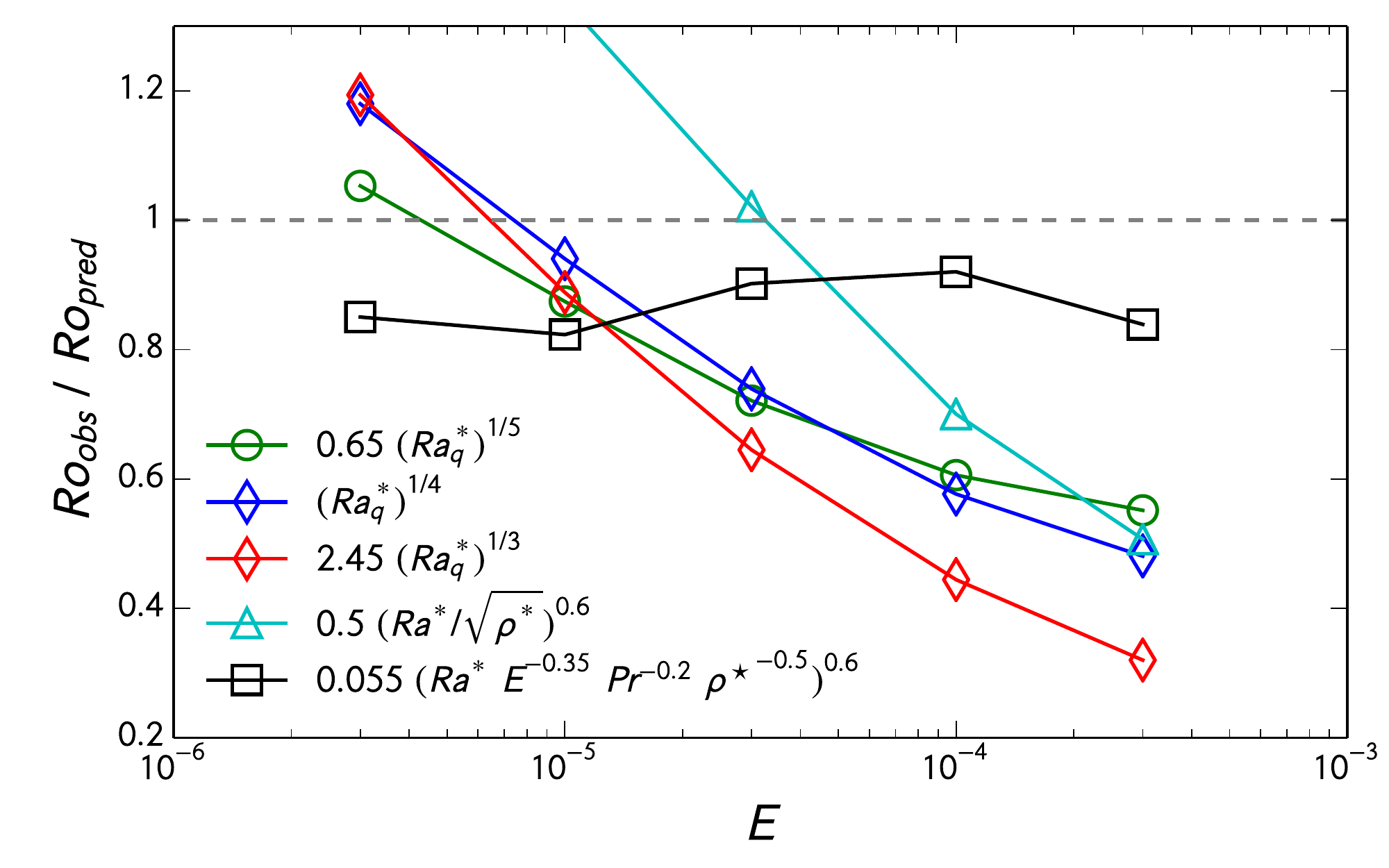}
 \caption{\modif{Test of the predictive power of different scaling laws: ratio 
of the actual zonal Rossby number ($Ro_{obs}$) of the different numerical 
models of Tab.~\ref{tab:runs} to the values predicted ($Ro_{pred}$) by 
different scaling laws. The horizontal dashed line corresponds to the ideal 
case of a perfect prediction ($Ro_{pred}=Ro_{obs}$).}}
 \label{fig:zfScaling}
\end{figure}

\modif{%
This scatter around the asymptotic laws therefore suggests that the 
zonal flow might obey a more complex scaling. The general form of a 
polynomial scaling for the zonal flow component can be written as 
the following function of the different control parameters:
\begin{equation}
 Ro_{\text{zon}} = 
\alpha\,{Ra^*}^\beta\,E^\delta\,Pr^\gamma\,{\rho^*}^\epsilon.
\end{equation}
A best-fit on the database of simulations then yields:
\begin{equation}
\begin{aligned}
 Ro_{\text{zon}} &= 0.055\left( Ra^*\,E^{-0.35}\,Pr^{-0.2}\,
{\rho^*}^{-0.5}\right)^{0.6}, \\
    &= 0.055\left( 
Ra\,E^{1.65}\,Pr^{-1.2}\,{\rho^*}^{-0.5}\right)^{0.6}.
\end{aligned}
\label{eq:zfScaling}
\end{equation}
This scaling has the same dependence on $Ra$ and $\rho^*$ as the 
previously obtained $Ra^*$ scaling but is not diffusivity-free
anymore owing to the additional dependence on $E$ and $Pr$. 
Fig.~\ref{fig:scaling}c shows that using this scaling significantly reduces the 
scatter. In addition, the anelastic numerical simulations of Tab.~\ref{tab:runs} 
(the five red stars) now fall very close to the asymptotic scaling law. This is 
encouraging as it means that Eq.~(\ref{eq:zfScaling}) correctly captures the 
dependence on the spherical shell mass and geometry.}

\begin{table*}
\caption{\modif{Table of results. The first column corresponds to the observed 
values for the mean surface zonal flow amplitude and the number of jets 
for the numerical simulation with $E=3\times 10^{-6}$, Jupiter and Saturn. The 
second, third and fourth columns correspond to the 
predicted values  using the convective-power scaling by \cite{Christensen02}, 
\cite{Showman10} and \cite{Gastine12}. The fifth column corresponds to the 
$Ra^*$ scaling, while the last column corresponds to the predicted values 
using the scaling given in Eq.~(\ref{eq:zfScaling}). Note that the zonal flow 
velocities for the numerical models have been obtained using the Saturnian 
rotation rate $\Omega$ and lengthscale $d$ (Tab.~\ref{tab:param}).}}
\centering
\begin{tabular}{ccccccc}
  \toprule
   & Obs. & $0.65\,(Ra_q^*)^{1/5}$ & $(Ra_q^*)^{1/4}$ & $2.45\,(Ra_q^*)^{1/3}$ 
& $0.5\,(Ra^*/\sqrt{\rho^*})^{0.6}$ & Eq.~(\ref{eq:zfScaling}) \\
  \midrule
  Zonal flow (m/s) & & & & & & \\
Simulation ($E=3\times 10^{-6}$) & 110 & 100 & 90 & 90 & 60 & 130 \\  
Jupiter & 45 & 3 & 2 & 1 & 0.3  &60\\
  Saturn  & 170 &  5 & 2 & 1 & 0.3 &100\\
  \midrule
  $n_{\text{jets}}$ & & & & & & \\
  Simulation ($E=3\times 10^{-6}$) & 12 & 10 & 11 & 11 & 13 & 9 \\
  Jupiter & 30 & 70 & 100 & 170  & 260 & 17 \\
  Saturn  & 17 & 50 & 80 & 150&  220 & 12\\
 \bottomrule
 \end{tabular}
\label{tab:results}
\end{table*}

\modif{%
Fig.~\ref{fig:zfScaling} shows a comparison of the predicting power of these 
different scaling laws. The ratio of the actual $Ro_{\text{zon}}$ of the 
numerical models to the predicted values are plotted for the five cases of 
Tab.~\ref{tab:runs} for the different aforementioned asymptotic scalings.
The power-based scalings  fail to predict the correct 
Rossby number with a better accuracy than 50\% for  $E \geq 10^{-4}$ and 
gradually approach the correct values at the lowest Ekman number. The 
$Ra^*$ scaling shows a similar trend at large Ekman number and even further 
deviates from the correct values at low Ekman numbers. In contrast, 
the scaling (\ref{eq:zfScaling}) predicts the Rossby number of 
the five numerical models within a 20\% error range without any notable 
remaining Ekman number dependence.}

\modif{Using the values given in Tabs.~\ref{tab:runs}-\ref{tab:param} allows 
to extrapolate these different scalings to Jupiter and Saturn, and
to evaluate the mean zonal flow amplitude and the number of jets. Confirming 
\cite{Showman10} and \cite{Gastine12}, the power-based scalings predict zonal 
flow velocities around 1-5~m/s for both Jupiter and Saturn and even smaller 
velocities are obtained when using the $Ra^*$ scaling. In contrast, 
our new scaling (\ref{eq:zfScaling}) predicts velocities in good agreement with 
the observed surface mean values. The number of jets are then derived using 
Eq.~(\ref{eq:jetscale}) and a mean $|\partial \ln M /\partial s|$ of $2\times 
10^{-7}$~m$^{-1}$ for Jupiter and $2.5\times 10^{-7}$~m$^{-1}$ for Saturn. The 
convective power scalings and the $Ra^*$ scaling overestimate the number of 
zonal bands by a factor of 3 to 5 for both planets. Although $n_{\text{jets}}$ 
is underestimated by roughly $1/2$ for both Saturn and Jupiter, 
the scaling (\ref{eq:zfScaling}) once again gives 
an improved prediction of the observed quantities.}

\modif{The scaling (\ref{eq:zfScaling}) obtained from a large database of 
Boussinesq and anelastic models is therefore encouraging as the predicted 
quantities are much closer to the observed values. It is nevertheless still a 
tentative scaling that would require additional theoretical work to further 
establish the obtained exponents on physical grounds.}

\section{Conclusion}
\label{sec:conclusion}

We have investigated several numerical simulations of rapidly-rotating 
convection in spherical shells. We have considered strongly stratified 
models with $\rho_\text{bot}/\rho_\text{top} \simeq 150$ in spherical 
shells with a large aspect ratio of $r_i/r_o=0.85$. Extending the previous 
Boussinesq study by HA07, we have explored a range of Ekman numbers 
($E=3\times 10^{-4}-3\times 10^{-6}$) to allow the formation of multiple high 
latitude jets. 

At the lowest Ekman number considered here, a strong prograde equatorial zonal 
flow is flanked by several alternating jets in each hemisphere. This confirms 
previous findings by \cite{Jones09} that compressible models can also maintain 
multiple zonal bands, provided low Ekman numbers and thin spherical shells are 
considered. The number of jets in our simulations is far fewer than observed 
for Jupiter. Smaller Ekman numbers, and even thinner shells would result 
in a greater number of narrower jets. However, both of these properties require 
computational expense that is presently impractical. Following our previous 
parameter studies of compressible convection \citep{Gastine12,Gastine13}, the 
zonal flow profiles obtained here show many similarities with their Boussinesq 
counterparts \citep[e.g.][HA07]{Heimpel05}. In agreement with the theory of 
geostrophic turbulence, the jet widths of these previous Boussinesq models have 
been found to scale with the Rhines length, i.e. $k_\beta^{-1} = 
\sqrt{2U/\beta}$, $\beta$ being the relative variation of the height of the 
spherical shell with the distance to the rotation axis \citep{Rhines75}. The 
similarities in the zonal flow profiles of Boussinesq and anelastic models 
would therefore suggest that this topographic Rhines scale could be also 
applicable to compressible convection. This seems however at odds with the 
two-dimensional anelastic models by \cite{Evonuk08} and \cite{Glatz09} which 
suggested a possible modification of the Rhines lengthscale due to 
compressibility effects.

To address this issue, we have derived here the expression of the 
$\beta$-effect in a 3-D spherical shell filled with a compressible fluid. 
Confirming the calculations by \cite{Ingersoll82}, the compressional $ 
\beta$-effect (i.e. $\beta_\rho$) is found to be proportional to the relative 
variation of the integrated mass with the distance to the rotation axis. 
Similarly to the classical topographic Rhines scaling (i.e. $\beta_h$), the 
compressional $\beta$-effect marks a sign reversal at the tangent cylinder. In 
the thin spherical shells considered here, the $\beta_\rho$ and $\beta_h$ 
profiles are found to be very similar inside the tangent cylinder. This 
explains the similarities of the high-latitude jets observed in Boussinesq and 
anelastic models. The main differences between topographic and compressional 
$\beta$-effects occur outside the tangent cylinder where the density gradient is 
mainly perpendicular to the rotation axis. However the Rhines scaling does not 
seem to be applicable there as there is only one single equatorial jet which 
seems to be always controlled by the geometry of the spherical shell (see also 
HA07). An efficient test of the compressional Rhines scale for $s> r_i$ would 
therefore require multiple zonal bands outside the tangent cylinder. While those 
low-latitude jets are frequently observed in the transient phases of the 3-D 
numerical models, they always merge on longer time resulting in one single 
prograde equatorial flow flanked by a retrograde jet attached to the tangent 
cylinder \citep[e.g.][]{Sun95,Christensen01}. Although 
numerically challenging,  3-D models in thicker spherical shells (i.e. 
$r_i/r_o \leq 0.6$) at low Ekman numbers (i.e. $E \sim 10^{-6}$) may help to 
check the stability of the zonal winds outside the tangent cylinder.
As the differences between $\beta_h$ and $\beta_\rho$ are also more pronounced 
in thicker shells, it would be interesting to check the applicability of the 
Rhines scaling to these low-latitude jets.

The compressional $\beta$-effect has also been applied to Jupiter's and 
Saturn's surface winds. For Jupiter, $\beta_h$ and $\beta_\rho$ are relatively 
similar and both predict jet-widths in relatively good agreement with the 
observed values. For Saturn, the misfit between predictions and observations is 
on average larger. The compressional Rhines scale gives nevertheless a 
better estimate of the jet-width than the topographic one. Excluding the 
equatorial jet, the best fit is achieved for a range of radius 
ratios of $\eta\simeq 0.85-0.9$ for Jupiter and $\eta\simeq 0.55$ for 
Saturn. These values are on the low-side compared to the estimates derived by 
\cite{Liu08}. However these depths are relatively poorly constrained by the 
Rhines scaling owing to the small changes when the radius ratio 
is varied. In addition, the factor 2 entering the Rhines scale definition is 
not firmly established as it comes from an \textit{ad-hoc} averaging of the 
orientation of the Rossby waves \citep{Rhines75}. \cite{Suko07} therefore argues 
that this factor should be replaced by a rather elusive coefficient of 
proportionality. It is therefore difficult to put a strong 
constraint on the zonal flow depth based on $\beta_\rho$ only.
\modif{Using the compressional Rhines scale also allows to construct an 
asymptotic scaling for the number of zonal jets. However, estimating this 
number requires to employ an asymptotic scaling for the zonal flow amplitude. 
The disagreement between the different buoyancy power scalings published to 
date suggests that the zonal flow might obey a more complex scaling. 
Thus, we propose a simple best-fit scaling law that successfully predicts 
the correct zonal flow amplitude and number of bands for both the numerical 
models and the gas giant planets. Although encouraging, the predicted 
dependence on the control parameters $Ra^*$, $E$ and $Pr$ is still tentative and 
needs further theoretical investigations to confirm and justify these 
exponents.}

\modif{The exact structure of the zonal jets in the gas giant interiors is 
still a matter of debate. The anelastic models by \cite{Kaspi09} for instance 
suggest an important variation of the zonal flow amplitude along the direction 
of the rotation axis that is not captured by our quasi-geostrophic zonal flow 
profiles. Testing the applicability of the compressional Rhines scaling to this 
kind of baroclinic jets is therefore an interesting prospect that would need to 
be addressed by future models.}

Finally a deeper understanding of the jet scaling would also require the study 
of integrated models that include the deep-interior dynamics.
Recent Boussinesq and anelastic dynamo models with radially decreasing 
electrical conductivity have shown that the penetration
of the zonal flows is limited by Lorentz forces that increase
strongly at depth \citep{Stanley09,Heimpel11,Duarte13}. While those models 
still show a strong prograde equatorial zonal flow, they lack well-defined 
high-latitude jets. Further investigations are therefore needed to analyse the 
depth and the stability of these high-latitude zonal winds.

\section*{Acknowledgements}

We thank P.~L.~Read for providing us Saturn's surface zonal wind profiles.
All the computations have been carried out on WestGrid, a regional 
division of Compute Canada. TG is supported by the Special Priority Program 
1488 (PlanetMag, \url{www.planetmag.de}) of the German Science 
Foundation. MH is supported in part by an NSERC Discovery Grant.
All the figures have been generated using matplotlib 
\citep[\url{www.matplotlib.org}, see][]{matplotlib}. 


\bibliographystyle{model2-names}

\end{document}